\documentclass[%
 aip,
 amsmath,amssymb,
 reprint,%
]{revtex4-1}

\usepackage{graphicx}
\usepackage{dcolumn}
\usepackage{bm}

\usepackage[utf8]{inputenc}
\usepackage[T1]{fontenc}
\usepackage{mathptmx}
\usepackage{etoolbox}
\usepackage{xcolor}

\makeatletter
\def\@email#1#2{%
 \endgroup
 \patchcmd{\titleblock@produce}
  {\frontmatter@RRAPformat}
  {\frontmatter@RRAPformat{\produce@RRAP{*#1\href{mailto:#2}{#2}}}\frontmatter@RRAPformat}
  {}{}
}%
\makeatother
\begin{document}

\preprint{AIP/123-QED}

\title[Harnessing Complexity]{Harnessing Complexity: \\Nonlinear Optical Phenomena in L-Shapes, Nanocrescents, and Split-Ring Resonators}
\author{Michael R. Clark}
\affiliation{Department of Physics, Arizona State University, Tempe, AZ 85287, USA}
\affiliation{Center for Nonlinear Studies (CNLS), Theoretical Division, Los Alamos National Laboratory, Los Alamos, NM 87545, USA}
\author{Syed A. Shah}
\affiliation{Center for Nonlinear Studies (CNLS), Theoretical Division, Los Alamos National Laboratory, Los Alamos, NM 87545, USA}
\author{Andrei Piryatinski}
\affiliation{Theoretical Division, Los Alamos National Laboratory, Los Alamos, NM 87545, USA}

\author{Maxim Sukharev}
\email[]{maxim.sukharev@asu.edu}
\homepage[]{http://sukharev.faculty.asu.edu}
\affiliation{Department of Physics, Arizona State University, Tempe, AZ 85287, USA}
\affiliation{College of Integrative Sciences and Arts, Arizona State University, Mesa, Arizona 85212, USA}

\date{\today}

\begin{abstract}
{We conduct systematic studies of the optical characteristics of plasmonic nanoparticles that exhibit C$_{2v}$ symmetry. Specifically, we analyze three distinct geometric configurations: an L-type shape, a crescent, and a split-ring resonator shaped like the Greek letter $\pi$. Optical properties are examined using the finite-difference time-domain (FDTD) method. It is demonstrated that all three shapes exhibit two prominent plasmon modes associated with the two axes of symmetry. This is in addition to a wide range of resonances observed at high frequencies corresponding to quadrupole modes and peaks due to sharp corners. Next, to facilitate nonlinear analysis, we employ a semiclassical hydrodynamic model where the electron pressure term is explicitly accounted for. This model goes beyond the standard Drude description and enables capturing nonlocal and nonlinear effects. Employing this model enables us to rigorously examine the second-order angular resolved nonlinear optical response of these nanoparticles in each of the three configurations. Two pumping regimes are considered, namely continuous wave (CW) and pulsed excitations. For CW pumping, we explore the properties of the second harmonic generation (SHG). Polarization and angle-resolved SHG spectra are obtained, revealing strong dependence on the nanoparticle geometry and incident wave polarization. The C$_{2v}$ symmetry is shown to play a key role in determining the polarization states and selection rules of the SHG signal. For pulsed excitations, we discuss the phenomenon of broadband terahertz (THz) generation induced by the difference-frequency generation (DFG). It is shown that the THz emission spectra exhibit unique features attributed to the plasmonic resonances and symmetry of the nanoparticles. The polarization of the generated THz waves is also examined, revealing interesting patterns tied to the nanoparticle geometry. To gain deeper insight, we propose an analytical theory that agrees very well with the numerical experiments. The theory shows that the physical origin of the THz radiation is the mixing of various frequency components of the fundamental pulse by the second-order nonlinear susceptibility. An expression for the far-field THz intensity is derived in terms of the incident pulse parameters and the nonlinear response tensor of the nanoparticle. The results presented in this work offer new insights into the linear and nonlinear optical properties of nanoparticles with C$_{2v}$ symmetry. The demonstrated strong SHG response and efficient broadband THz generation hold great promise for applications in nonlinear spectroscopy, nanophotonics, and optoelectronics. The proposed theoretical framework also provides a valuable tool for understanding and predicting the nonlinear behavior of other related nanostructures.}
\end{abstract}

\maketitle

\begin{quotation}
{}
\end{quotation}

\section{Introduction}
The unique optical properties of noble metals in the visible part of the spectrum have captivated scientists and artists for centuries.\cite{Song2019} Dating back hundreds of years, the distinctive scattering of small gold particles, which depend sensitively on their sizes and shapes, was cleverly exploited in the creation of stained glass. This early recognition of the fascinating interplay between light and metallic nanostructures laid the foundation for the modern research field of nano-plasmonics,\cite{Wang2011} which has now permeated the realms of physics,\cite{Zhang2012} chemistry,\cite{Odom2011} engineering,\cite{Maier2001} and materials science.\cite{Barnes2007} The rapid advancement of this field can be attributed to the tremendous progress in fabrication techniques\cite{annurev_fabrication} and laser technologies, which have enabled researchers to explore the intriguing phenomena arising from the interaction of light with metallic nanostructures at an unprecedented level of detail.

One of the most exciting features of plasmonic materials is their ability to strongly concentrate light well beyond the diffraction limit when driven at a corresponding surface-plasmon polariton (SPP) frequency. This remarkable property stems from the collective oscillation of conduction electrons confined within the metal, which gives rise to a strong field localization at metal-dielectric interfaces. The experimental capabilities to fabricate virtually any imaginable\cite{Chen2015,Urban2019} (and sometimes even unimaginable\cite{Malkiel2018}) interfaces with outstanding precision, coupled with the strong dependencies of SPP modes on shapes and light polarization, have rendered the research field of nano-plasmonics highly applicable across a wide range of scientific and technological disciplines.

The remarkable optical properties observed in plasmonic systems operating within the linear regime have sparked a renewed interest in utilizing plasmons as a foundation for future nonlinear nano-devices. The linear regime has traditionally been the primary focus of plasmonics research.\cite{Maier2001} Nevertheless, recent advancements have unveiled the potential of plasmonic systems to be employed as a platform for nonlinear optics, which involves the study of light-matter interactions in highly intense fields. Metals have high nonlinear susceptibilities but usually high Ohmic losses limit utilization of such systems in nonlinear devices as the latter require macroscopic propagation lengths. However, never-ending quest for miniaturizing optical devices has also led to realization that the ability of plasmonic systems for strong light localization under SPP resonant conditions\cite{Gramotnev2010} can be utilized to explore a wide range of nonlinear optical processes at true nanoscale.\cite{Panoiu2018}  By harnessing the nonlinear properties of plasmons, it is possible to manipulate and control light in ways that were previously unattainable. This has opened the door to the development of innovative nonlinear devices that can go beyond the limitations of traditional linear optical systems. Consequently, the investigation of plasmonics in the nonlinear regime has gained considerable attention from researchers and scientists alike, offering a promising avenue for the advancement of nonlinear optics and the realization of novel applications in various fields.

Clearly the field of plasmonics has made notable strides in recent years, particularly in the context of engineering applications. However, its implications extend far beyond, notably impacting the realm of polaritonic chemistry.\cite{Fregoni2022} This emerging interdisciplinary area leverages the small volume occupied by plasmon modes to facilitate strong coupling interactions between molecular excitons and plasmons. References such as Torma et al.\cite{TormaReview} and Sukharev et al.\cite{Sukharev2017} have extensively reviewed these developments, emphasizing their potential in advancing our understanding of light-matter interactions at the nanoscale.

Recent theoretical and experimental investigations have further expanded our understanding of these interactions. For instance, studies by Sukharev et al. in 2018\cite{Sukharev2018} and Drobnyh et al.\cite{Drobnyh2019} in 2019 provided theoretical frameworks that support the observation of polaritonic states, which emerge when excitons and plasmons strongly couple. These states have been directly observed in experiments such as those reported by Li et al. in 2020,\cite{Li2020} which detailed their manifestation in phenomena like second harmonic generation. This particular nonlinear optical process has proven to be a valuable tool in probing the unique properties of polaritonic states.

Further investigations have revealed that the optical properties of these coupled states can be finely tuned by adjusting the pumping conditions, enabling controlled directional optical transitions between them.\cite{Sukharev_Salomon_Zyss_2021} Such control offers promising avenues for the development of novel optical devices and sensors.

Moreover, the unique symmetry properties inherent to second harmonic generation have been exploited in recent experimental endeavors, such as those using gold nanocrescents.\cite{Maekawa2020} These studies demonstrate the potential of this technique to detect minor spatial deviations in nanostructures, which could be crucial for applications requiring high precision in nanoparticle fabrication and deployment. This capability not only underscores the sensitivity of plasmonic structures to shape and size but also highlights the broader applicative potential of second harmonic generation in materials diagnostics and characterization.

In this work, we focus on periodic arrays comprised of nanoparticles with three distinct shapes characterized by the C$_{2v}$ symmetry group. By employing the semiclassical hydrodynamic model,\cite{Eguiluz1976} we investigate the second harmonic generation (SHG) and explore the difference-frequency generation (DFG) by short femtosecond pulses resulting in highly coherent bursts of terahertz (THz) radiation.\cite{Fang2018} The latter exhibits various intriguing properties that hold great promise for future THz  applications. Our findings contribute to the growing body of knowledge in the field of nonlinear nano-plasmonics and provide valuable insights into the design and optimization of novel nanoscale nonlinear optical devices. 

The manuscript is organized as follows: our theoretical model is discussed in Sec.~\ref{sec:model}, Sec.~\ref{Sec:SimDepict} provides parameters of our numerical simulations, the main results for linear optical properties and SHG of the geometries considered are provided in Sec.~\ref{Sec:LinearSHG}, the discussion of the THz broadband radiation generated by fs pulsed excitation is provided in Sec.~\ref{Sec:THzPulseSim}, analysis of the signal polarization dependence and spectral lineshape utilizing an analytical model is given in Sec.~\ref{Sec:TheoryLinSh}. In Sec.~\ref{Sec:conclusion}, we summarize our main findings and provide an outlook.

\section{Theoretical model\label{sec:model}}
We consider three shapes of Au nanoparticles depicted in Fig.~\ref{fig:Figure1}, namely the L-Shape particle (LS), the Split Ring Resonator (SRR), and the Nanocrescent particle (NC). These particles are arranged in a square two-dimensional lattice with a period of 400 nm. The geometries are oriented within the $(x,y)$ laboratory axes, which describe the periodicity of each lattice. A second coordinate system, $(u,v)$, is defined as the $u$-axis always oriented along the nanoparticle symmetry axis. As shown below, the polarization of plasmonic eigenmodes naturally aligns with the $(u,v)$-coordinate system. Thus, we refer to this coordinate system as the natural system throughout this paper. Each geometry is irradiated by an incident linearly-polarized electromagnetic field, with the electric component polarized within the $(x,y)$-plane and forming an angle $\theta$ with respect to the $x$-axis. The field propagates perpendicular to the plane in the negative direction of the $z$-axis.

The strength of the incident field is set to either emphasize the nonlinear response of the system (with an electric field strength on the order of $10^7$ V/m) or is set low enough such that the nonlinear response is negligible (on the order of 1 V/m). The goal of the low strength incident field is to scan through a range of frequencies (converted into energies on the order of eV) to determine the resonant frequencies / energies of the system, as shown in Figure~\ref{fig:Figure1}. Then we simulate the system with the higher incident field strength pumped at the resonant energies, to increase the nonlinear optical response of the system.

To model the light propagation through the array, we start by numerically solving Maxwell's equations
\begin{eqnarray}
    \nonumber
    \dot{\bm{B}} &=& - \bm{\nabla}\times\bm{E}, \\
    \epsilon_0\dot{\bm{E}} &=&\frac{1}{\mu_0}\bm{\nabla}\times\bm{B}-\dot{\bm{P}},
    \label{eq:Maxwell}
\end{eqnarray}
where the macroscopic polarization $\bm{P}$ satisfies the equation of motion following from the semi-classical hydrodynamics model\cite{Scalora_Vincenti_de_Ceglia_Roppo_Centini_Akozbek_Bloemer_2010}. We note that from the numerical point of view it is easier to deal with the second-order differential equation material containing polarization\cite{Drobnyh2020} rather than the equation depending on the current density, $\bm{J}$. The equation determining the dynamics of conduction electrons reads
\begin{eqnarray}
    \nonumber
    \ddot{\bm{P}} + \gamma\dot{\bm{P}} = \dfrac{n_0e^2}{m^*}\bm{E}-\dfrac{e}{m^*}\left(\bm{\nabla}\cdot\bm{P}\right)\bm{E} + \dfrac{e}{m^*c}\dot{\bm{P}}\times\bm{H} \\
     - \dfrac{1}{n_0e}\left[\left(\bm{\nabla}\cdot\dot{\bm{P}}\right)\dot{\bm{P}} + \left(\dot{\bm{P}}\cdot\bm{\nabla}\right)\dot{\bm{P}}\right] - \dfrac{e\nabla{}p}{m^*},
    \label{eq:Hydrodynamic}
\end{eqnarray}
where $n_0$ is the electron density at equilibrium, $m^*$ is the effective mass of conduction electrons, and $\nabla{}p$ is the phenomenological quantum electron pressure term. We determine the electron density by considering the integral of the continuity equation $\int dt\dot{n}=-\int dt \frac{1}{e}\bm{\nabla}\cdot\dot{\bm{P}} = n_0 - \frac{1}{e}\bm{\nabla}\cdot\bm{P}$. 

The term describing the electron pressure is frequently depicted as $p=n_0E_F(n/n_0)^{5/3}$, where $E_F$ is the Fermi energy\cite{Scalora_Vincenti_de_Ceglia_Roppo_Centini_Akozbek_Bloemer_2010,Linder_1936,Manfredi_2001}. $\nabla{}p$ then becomes  
\begin{eqnarray}
    \nabla p = \frac{5}{3}\frac{n_0E_F}{n_0^{5/3}}n^{2/3}\nabla{n}.
    \label{eq:E-PressureStep}
\end{eqnarray}

Using the above relationship between $n$ and $\bm{P}$, we substitute $n$ for $\bm{P}$ and expand the resulting expression up to the third order,
\begin{eqnarray}
    \nonumber
    \nabla{}p = \dfrac{5}{3}\dfrac{E_F}{e}\nabla\left(\bm{\nabla}\cdot\bm{P}\right) - \dfrac{10}{9}\dfrac{E_F}{n_0e^2}\left(\bm{\nabla}\cdot\bm{P}\right)\nabla\left(\bm{\nabla}\cdot\bm{P}\right) \\
    - \dfrac{5}{27}\dfrac{E_F}{n_0^2e^3}\left(\bm{\nabla}\cdot\bm{P}\right)^2\nabla\left(\bm{\nabla}\cdot\bm{P}\right).
    \label{eq:E-Pressure}
\end{eqnarray}

We note that Eq.~\eqref{eq:E-Pressure} contains the linear term, which happens to play an important role for particles with sharp corners. The coupled Eqs.~\eqref{eq:Maxwell}, \eqref{eq:Hydrodynamic}, and \eqref{eq:E-Pressure} constitute our model used to scrutinize both linear and nonlinear optical properties of plasmonic systems. 

The following parameters are used to simulate gold: $n_0=5.9\times10^{28}$ m$^{-3}$, $m^* = 1.66\times{}m_e$, $\gamma_e=0.181$ eV, and $E_F = 5.53$ eV. 

\section{Numerical procedure}
\label{Sec:SimDepict}
The system is discretized in both space (through the $(x,y)$-axes) and time using the finite-difference time-domain (FDTD) methodology\cite{Taflove_Hagness_2010} with spatial resolution $1.5$ nm in all three dimensions and the time step of $2.5$ as. The simulated system size is 396 nm$\times$396 nm$\times$1440 nm, resulting in a total of $6.7\times10^7$ grid points for each of the three separate coordinate systems which are offset in one of the three dimensions in accordance with the FDTD methodology. Periodic boundary conditions are imposed along the $x$ and $y$ axes and absorbing boundary conditions (implementing using the convolutional perfectly matched layers methodology\cite{Gedney_2001}) are imposed along the $z$-axis, to absorb outgoing radiation.

Additionally, a piecewise window function based on the Blackman-Harris time window \cite{Aguirregabiria_2017} is applied to the detected electric field components to eliminate spurious contributions that occur due to the spatial discretization of the system. The time window uses the formula for $t\leq\tau$
\begin{eqnarray}
    a_0 + a_1\cos{\frac{2\pi{}t}{\tau}} + a_2\cos{\frac{4\pi{}t}{\tau}} + a_3\cos{\frac{6\pi{}t}{\tau}},
    \label{Blackman-Harris}
\end{eqnarray}
where $\tau$ is the propagation time of the incident field, and $a_{(0,1,2,3)} = (0.3532, -0.488, 0.145, -0.0102)$. For time $t>\tau$ but less than the total propagation time ($t_{\text{final}}$) of the simulated system, the function is set to zero. This time window is incredibly successful in drastically reducing the noise in the nonlinear spectra, as seen when comparing Figs.~\ref{fig:Figure3}a and \ref{fig:Figure5a}a. The reason for the removal of the time window is to see other resonant peaks in the nonlinear spectra, specifically the THz peak, which is explored in more detail in Section \ref{Sec:THzPulseSim}.

The home-built codes utilize Message Passing Interface (MPI) for efficient parallel simulations, dividing the system into 1152 sub-domains with send/receive operations applied to all six faces of each sub-domain. Simulations are performed on Los Alamos National Laboratory's HPC cluster Chicoma, a HPE Cray EX system. Typical execution times depend on whether the nonlinear or linear spectra is simulated, with a range between several minutes to about two hours for the most intensive simulations.

The total propagation time is set to $t_{\text{final}}= 400$ fs for the linear response and $t_{\text{final}}= 500$ fs for the nonlinear response. We note that, in general, linear spectra for plasmonic systems in the visible can be obtained with a good precision setting $t_{\text{final}}$ to a lifetime of a plasmon mode of interest. Usually this is well below $400$ fs. However, one needs to insure numerical convergence especially for geometries containing sharp corners. We thus tested our simulations and found that $400$ fs propagating time of linear equations results in converged spectra. Similarly, the nonlinear simulations have been tested with regards to obtaining converged lineshapes of both SHG and DFG signals.
In both the linear and nonlinear simulations the polarization of the incident electric field is given by the angle $\theta$ with respect to the $x$-axis. 

\section{Linear response and the second harmonic generation}
\label{Sec:LinearSHG}

\begin{figure*}[ht]
\centering
\includegraphics[width=0.8\textwidth]{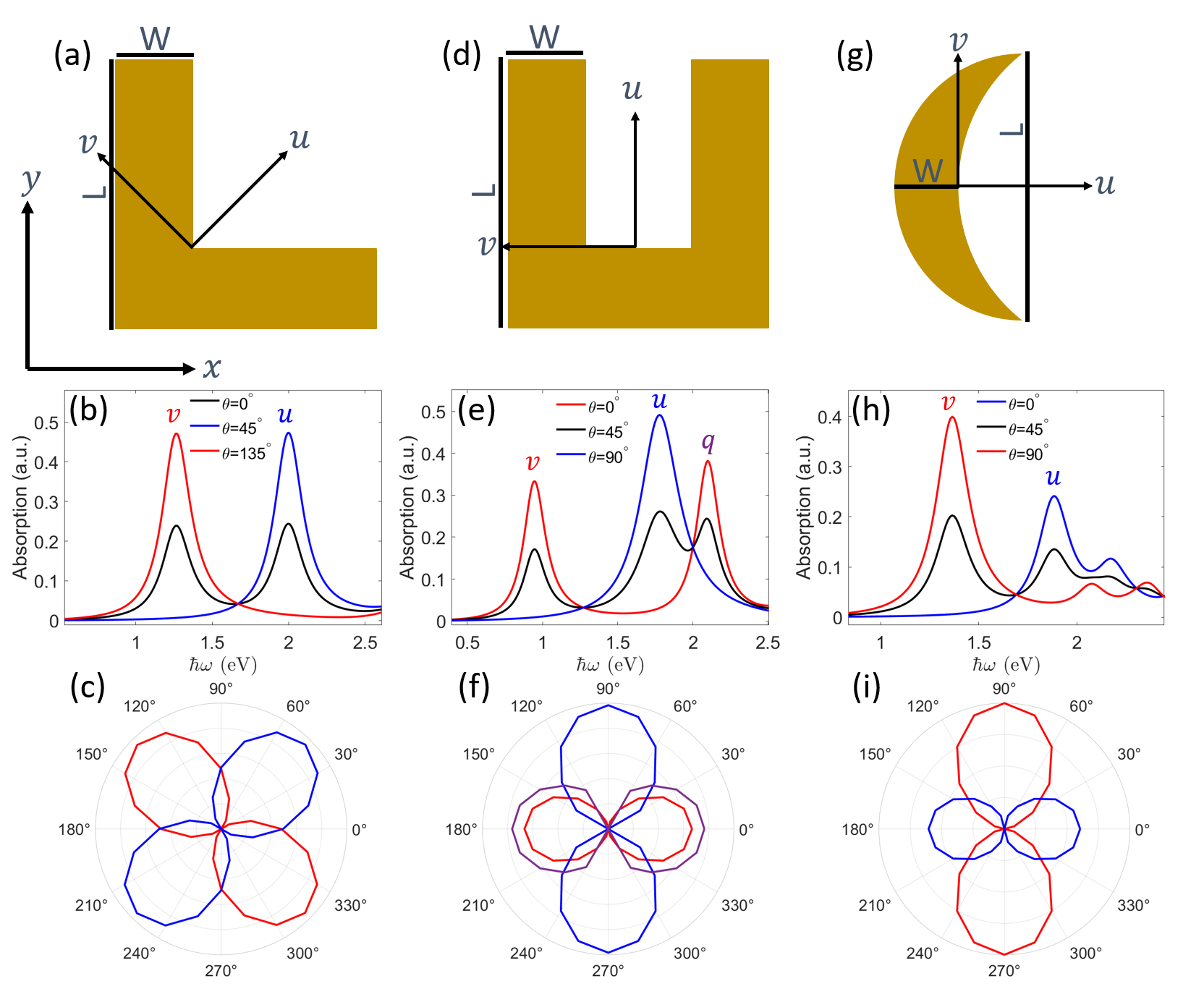}
\caption{(a,d,g) The three geometries as plotted in the $xy$-plane. In all three geometries $L=130$ nm and $W=40$ nm with each nanoparticle being $30$ nm thick. They are oriented with respect towards the laboratory axes $(x,y)$, with their natural axes $(u,v)$ shown on the geometry itself. (b,e,h) Each geometries' linear absorption as a function of the incident photon energy ($\hbar\omega$). Black, blue, and red colors in each panel correspond to the incident field polarization - (b) black shows $\theta=0^{\circ}$, blue is for $\theta=45^{\circ}$, and red is for $\theta=135^{\circ}$, (e) red shows $\theta=0^{\circ}$, black is for $\theta=45^{\circ}$, and blue is for $\theta=90^{\circ}$, (h) blue shows $\theta=0^{\circ}$, black is for $\theta=45^{\circ}$, and red is for $\theta=90^{\circ}$. (c,f,i) Each geometries' resonant absorption peaks as functions of the initial E-field polarization $\theta$. (a,b,c) The LS has two resonant peaks in its linear spectra, at $\hbar\omega_v=1.2614$ eV, where $\theta = 135^{\circ}$ (shown in red) and at $\hbar\omega_u=1.9955$ eV, where $\theta = 45^{\circ}$ (shown in blue). (d,e,f) The SRR has three resonant peaks, at $\hbar\omega_{v}=0.9512$ eV where $\theta = 0^{\circ}$ (shown in red), at $\hbar\omega_{u}=1.7783$ eV where $\theta = 90^{\circ}$ (shown in blue), and at $\hbar\omega_{q}=2.0989$ eV where $\theta = 0^{\circ}$ (also shown in red, but with a deep purple $q$ marking the 3rd resonance). (g,h,i) The NC has two resonant peaks, at $\hbar\omega_{v}=1.3648$ eV where $\theta =90^{\circ}$ (shown in red) and at $\hbar\omega_{u}=1.8817$ eV where $\theta =0^{\circ}$ (shown in blue).}
\label{fig:Figure1}
\end{figure*}

\begin{figure}[t]
\centering
\includegraphics[width=0.5\textwidth]{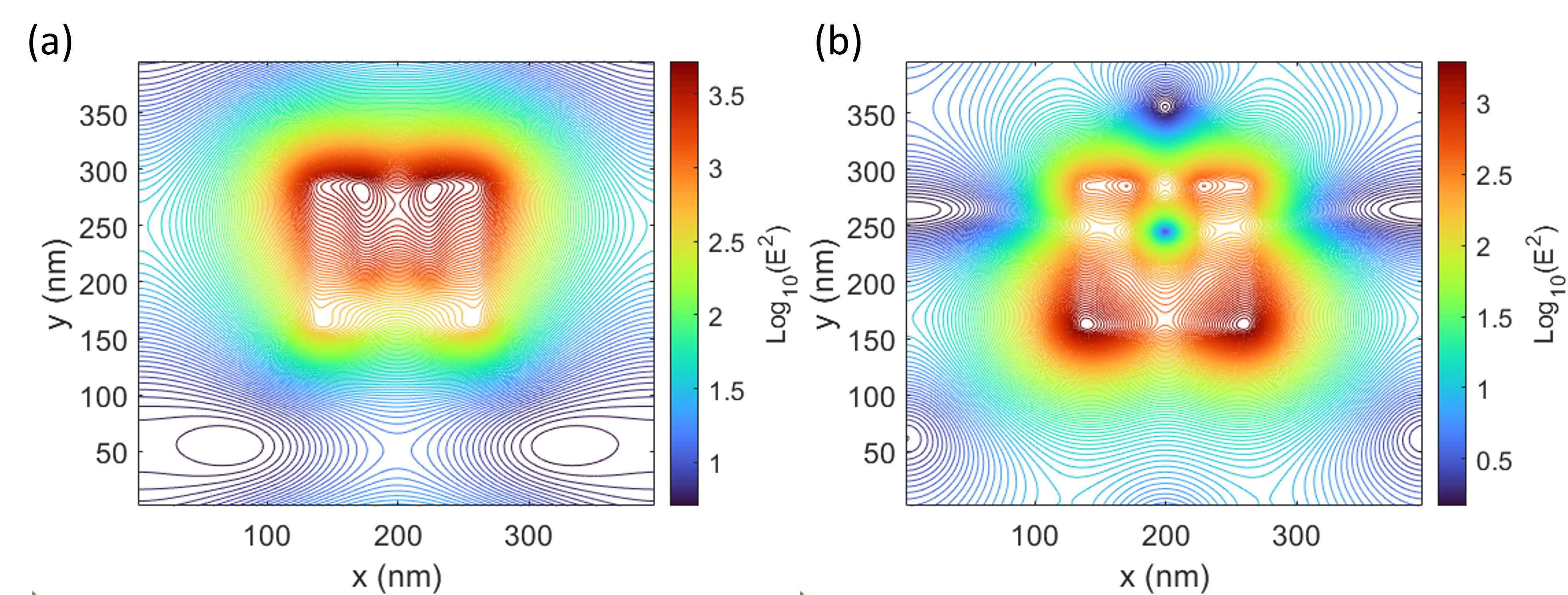}
\caption{A contour map of the logarithm of the electromagnetic field field intensity for the (a) $v$ and (b) $q$ SRR resonances. The detection plane is placed 15 nm above the top surface of the metallic nanoparticle. In these simulations, the incident electric field strength is 1 V/m. The units of intensity are normalized with respect to the incident field intensity in vacuum.}
\label{fig:Figure2}
\end{figure}

Fig.~\ref{fig:Figure1} shows a depiction of each geometry, including their size specifications. The differences between the LS and the SRR can be attributed to the inclusion of the second vertical arm to the Split Ring Resonator. This inclusion changes the resonant electric field polarization from $\theta=135^{\circ}$ to $\theta=0^{\circ}$ for the $v$ resonance (corresponding to the natural $v$-axis, and from $\theta = 45^{\circ}$ to $\theta=90^{\circ}$ for the $u$ resonance (corresponding to the natural $u$-axis), as well as the creation of a third resonant frequency, denoted $q$ as it does not constitute a third natural axis. Of note are the two SRR resonances along the $0^{\circ}$ polarization corresponding to the $v$ and $q$ resonances respectively, which are due to the dipole moment that occurs along the ends of the resonator's arms and surface plasmon-polaritons that travel along the base that connects the arms together respectively. This is shown in Fig.~\ref{fig:Figure2}.

\begin{figure}[t]
\centering
\includegraphics[width=0.5\textwidth]{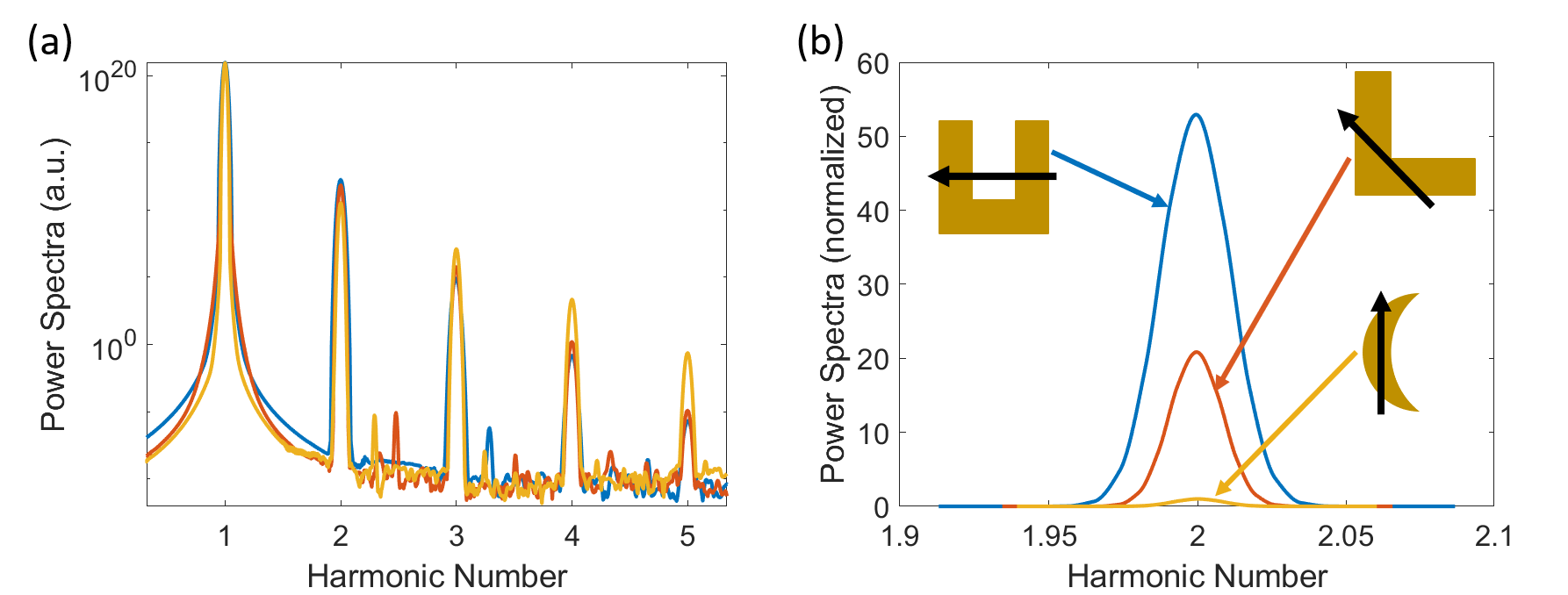}
\caption{(a) The strength of each geometries' nonlinear power spectra, corresponding to the angle of polarization for their $v$ resonance. Here, the blue line represents the power spectra of the SRR, the red represents the power spectra of the LS, and the yellow represents the power spectra of the NC. (b) A comparison of the SHG, normalized to the amplitude of the SHG of the NC.}
\label{fig:Figure3}
\end{figure}

Another aspect to note is the appearance of additional peaks at higher frequencies in Fig.~\ref{fig:Figure1}h, which can be attributed to the sharper end points of the NC, where SPPs can concentrate at higher densities, leading to stronger electric field intensities. This is why these peaks are only seen in the NC linear spectra, as its corners are significantly more sharp than the corners of the SRR and the LS. This effect can be seen more explicitly in Fig.~\ref{fig:Figure2}, where the fields are strongest at the corners. A way to reduce this concentration would be to round the corners of the geometries, which would spread out the SPPs. As can be seen, the fields are strongest along (a) the top demonstrating the dipole moment response and (b) the bottom depicting a bright mode along the base of the SRR that is directly excited by the incident field. 

The resonant frequencies $\omega_p$ obtained from the linear absorption spectra are used as electric field pump frequencies with magnitude $E_0$ which, when $E_0$ is set sufficiently strong, yields a nonlinear response in the gold particles. Fig.~\ref{fig:Figure3} shows the power spectrum for each geometry under intense illumination. We note the strongest peak for all three cases as expected corresponds to the elastic (Rayleigh) scattering. Both even and odd harmonics are observed. There is also noticeable noise that can be seen at intermediate frequencies between harmonics. Such a noise is actually significantly smaller compared to the case when one is not using the time window (see Eq.~\eqref{Blackman-Harris}). The latter clearly allows to compute not only relative strength of each harmonic but also their linewidths. 

\begin{figure*}[ht]
\centering
\includegraphics[width=0.8\textwidth]{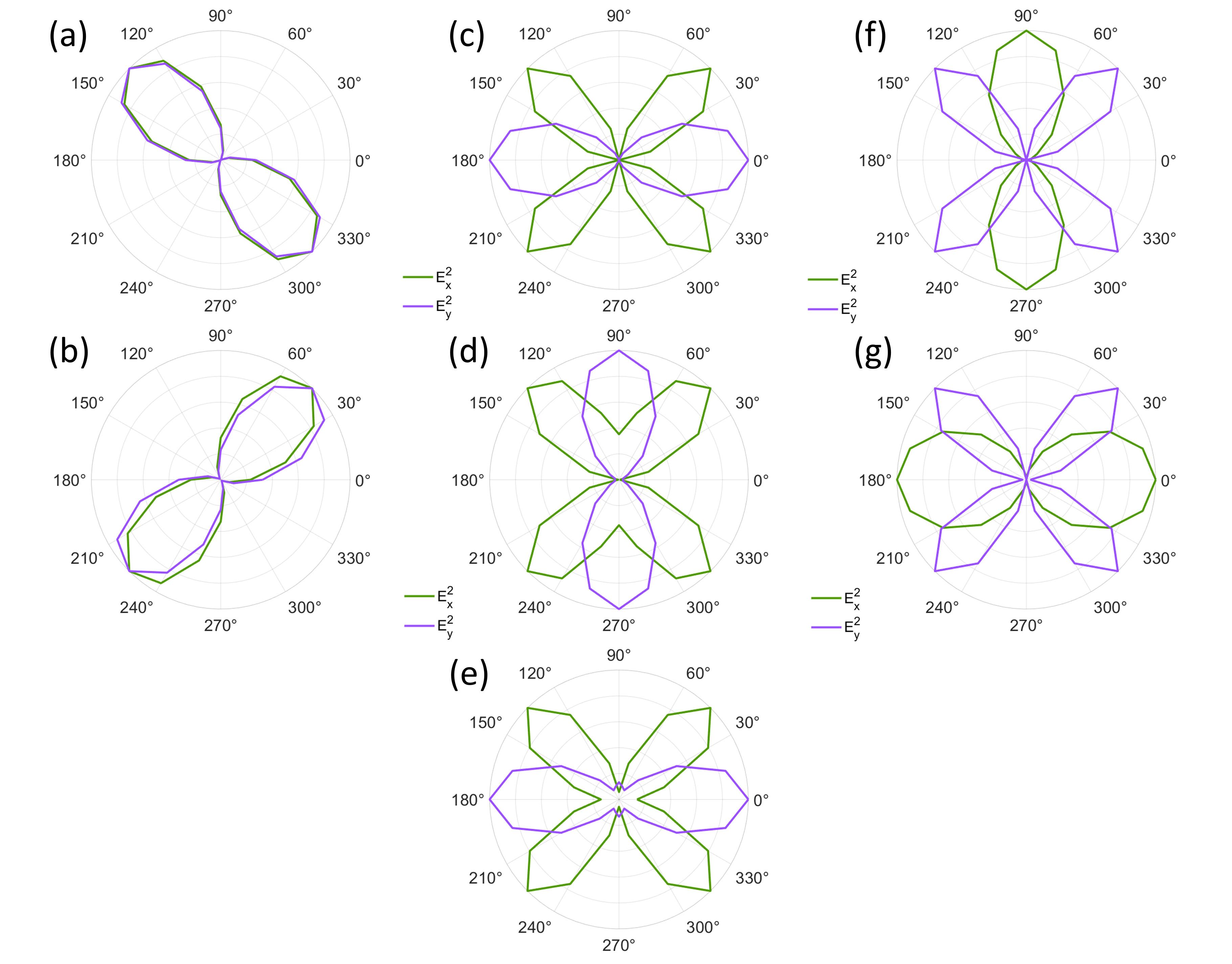}
\caption{The normalized SHG amplitude of the Fourier-transformed $E_x^2$ (shown in green) and $E_y^2$ (shown in purple) plotted as a function of the initial E-field polarization. (a,b) The LS x- and y-components for both $v$ and $u$ resonances. (c,d,e) The SRR x- and y-components for the $v$ (panel (c)), the $u$ (panel (d)), and the $q$ (panel (e)) resonances. (f,g) For the NC geometry, (f) shows results for the resonant $v$ frequency and (g) is for the pump at the resonant $u$ frequency.}
\label{fig:Figure4}
\end{figure*}

Fig.~\ref{fig:Figure3}b compares the strength of each geometries' SHG. The SRR, as expected\cite{Ciraci_2012}, has the strongest SHG response, followed by the LS with the NC coming in last. The SRR is 52 times the SHG of the NC, with the LS having less than half the strength of the SRR (about 21 times the SHG of the NC). Obviously, the strength of the SHG response does depend on the polarization of the incident electric field since non-zero components of the second-order susceptibility tensor, $\chi^{(2)}$, are not identical. Fig.~\ref{fig:Figure4} explores the angular dependence of the second harmonic intensity as a function of the incident field polarization. Here, each geometry is pumped at their respective resonant frequencies but with a varied angle of polarization $\theta$, and their normalized Fourier-transformed squared electric field components were extracted.

The $x$- and $y$-components for both the $v$ and $u$ resonances of the LS are approximately equal in magnitude (panels (a) and (b)). The $y$-component of the E-field for the SRR is overall the stronger component, being two orders of magnitude stronger when pumped at the $v$ resonance (panel (c)), approximately equal in magnitude at the $u$ resonance (panel (d)), and about an order of magnitude stronger at the $q$ resonance (panel (e)). The $x$-component in the NC is the overall stronger component, being about equal in magnitude when driven at the $v$ resonance (panel (f)) and being about two orders of magnitude stronger at the $u$ resonance (panel (g)).

The SHG responses from both SSR and NC show identical symmetry (Fig.~\ref{fig:Figure4} c-g);
however, the response for LS appears different (Fig.~\ref{fig:Figure4} a-b). 
This apparent difference is the consequence of $45^\circ$ rotation between the symmetry axis of LS and $x$, $y$ axis (as clarified in Sec.~\ref{Sec:PolDep}). The polarization corresponding to the strongest component of each of the resonances are in agreement with their corresponding linear spectra in the polar plots from Fig.~\ref{fig:Figure1}.

\section{TH\MakeLowercase{z} Pulse Generation}
\label{Sec:THzPulseSim}
One very intriguing feature not seen in Fig.~\ref{fig:Figure3} is the aforementioned THz pulse generation, which, as mentioned in Sec.~\ref{Sec:SimDepict}, is obscured in part due to the Blackman-Harris time window employed to reduce overall noise. Another reason for this peak not detected in Fig.~\ref{fig:Figure3} is the pump pulse duration as we explain below. Since the pulse occurs at very low energy ($<0.1$ eV) relative to the energy of the incident field, the pulse's amplitude, as well as its peak structure, might be obscured or affected in some way. Thus, in the following section the time window previously utilized has been removed, leading to Fourier-transformed electric field components that have a noticeably higher level of noise due to the spatial discretization of the system.

\begin{figure*}[ht]
\centering
\includegraphics[width=0.8\textwidth]{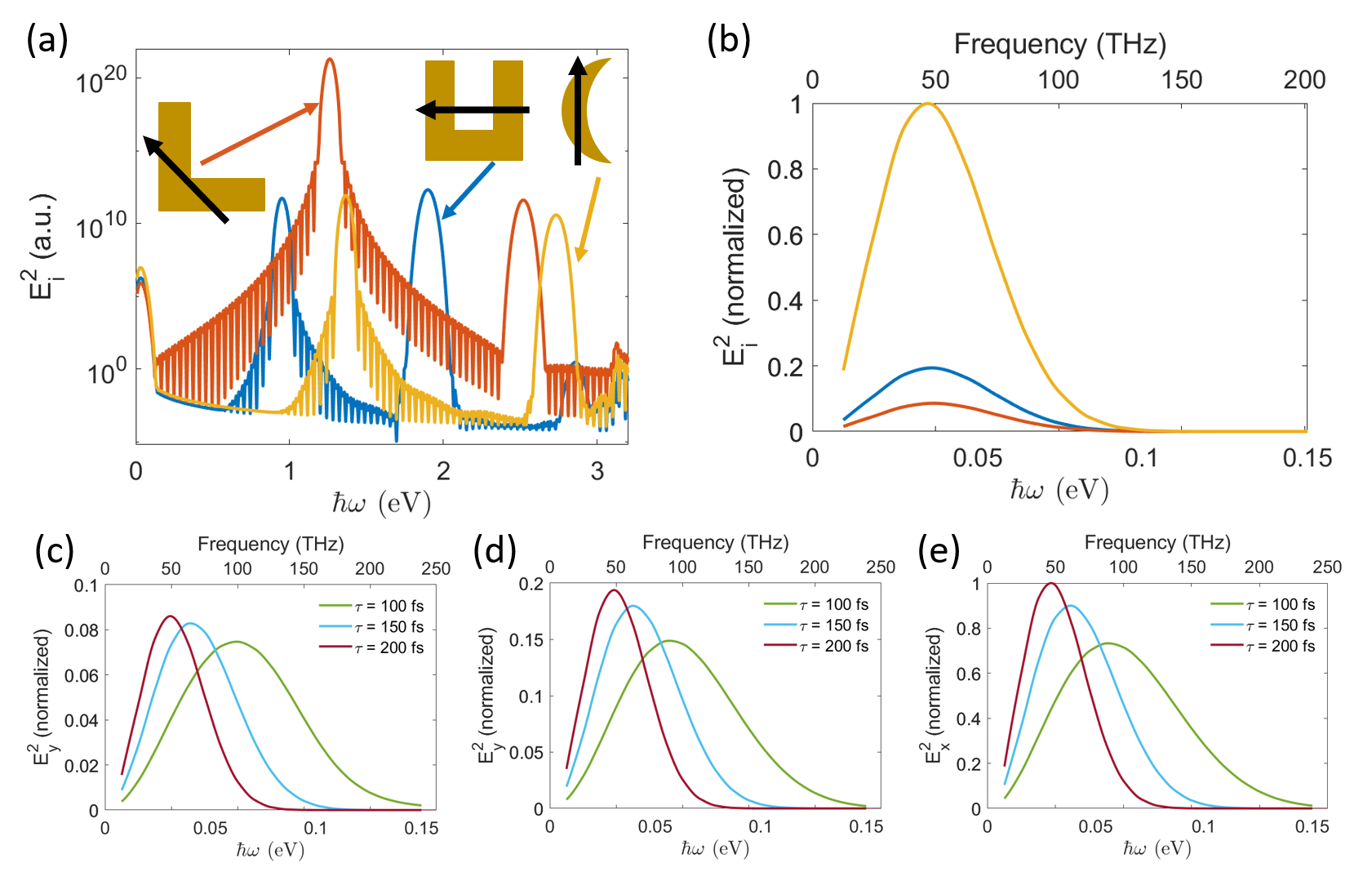}
\caption{The Fourier-transformed squared $x$- or $y$-component of the electric field as a function of the energy ($\hbar\omega$) with the incident electric field pump frequency and polarization set to the $v$ resonance for each geometry, where the SRR and LS geometries depict the $y$-component and the NC depicts the $x$-component. The incident pulse duration is $\tau = 200$ fs and the pump amplitude is $E_0=10^7$ V/m. The incident pulse polarization for each geometry is shown in the inset of panel (a). (a) Power spectra for all geometries a function of outgoing photon energy $\hbar\omega$. As in Fig.~\ref{fig:Figure3}, the SRR spectra are blue, the LS spectra are red, and the NC spectra are yellow. (b) Zooming into the $0.01 - 0.2$ eV shows the THz pulse, normalized by the amplitude of THz signal produced by the NC. (c,d,e) The THz peak for three different incident pulse durations ($\tau = $100 fs, 150 fs, and 200 fs) for the $u$ resonance of each geometry, with the same pattern as the above figures, that is, the LS is (c), the SRR is (d), and the NC is (e).}
\label{fig:Figure5a}
\end{figure*}

Fig.~\ref{fig:Figure5a} shows the spectra of the strongest Fourier-transformed electric field component for each geometry, as well as a zoomed in plot showing THz pulses normalized with respect to the THz signal observed for the NC when driven by the $200$ fs pulse. One important observation of Fig.~\ref{fig:Figure5a}b to note is the relative strength of the NC pulse compared to the other geometries; specifically, the NC is about five times stronger than the SRR pulse, and more than an order of magnitude stronger than the LS pulse, even though the LS spectra at its pump frequency is about ten orders of magnitude stronger than the NC at its pump frequency. This demonstrates a geometrical dependence on the strength of the THz pulse. However, while the geometry does not noticeably affect the frequency of the THz pulse, the pump pulse duration, $\tau$, does have a drastic effect on both its frequency, linewidth, and the strength, as seen in Figs.~\ref{fig:Figure5a}c, \ref{fig:Figure5a}d, and \ref{fig:Figure5a}e, where the pulse spreads out the shorter the propagation time is. The latter is the important observation explaining in part why longer propagation times (ideally leading to continuous wave excitation) do not result in THz signal observation. Numerically due to a finite frequency resolution THz peak is washed out moving to a non-zero signal at the frequency 0. The underling physical process leading to the observed THz peak in the power spectra is DFG caused by the finite bandwidth of the pump pulse.\cite{Fang2018} We expand the theory behind this in more details in the next section providing analytical expressions for the THz pulse characteristics.

\begin{figure*}[ht]
\centering
\includegraphics[width=0.8\textwidth]{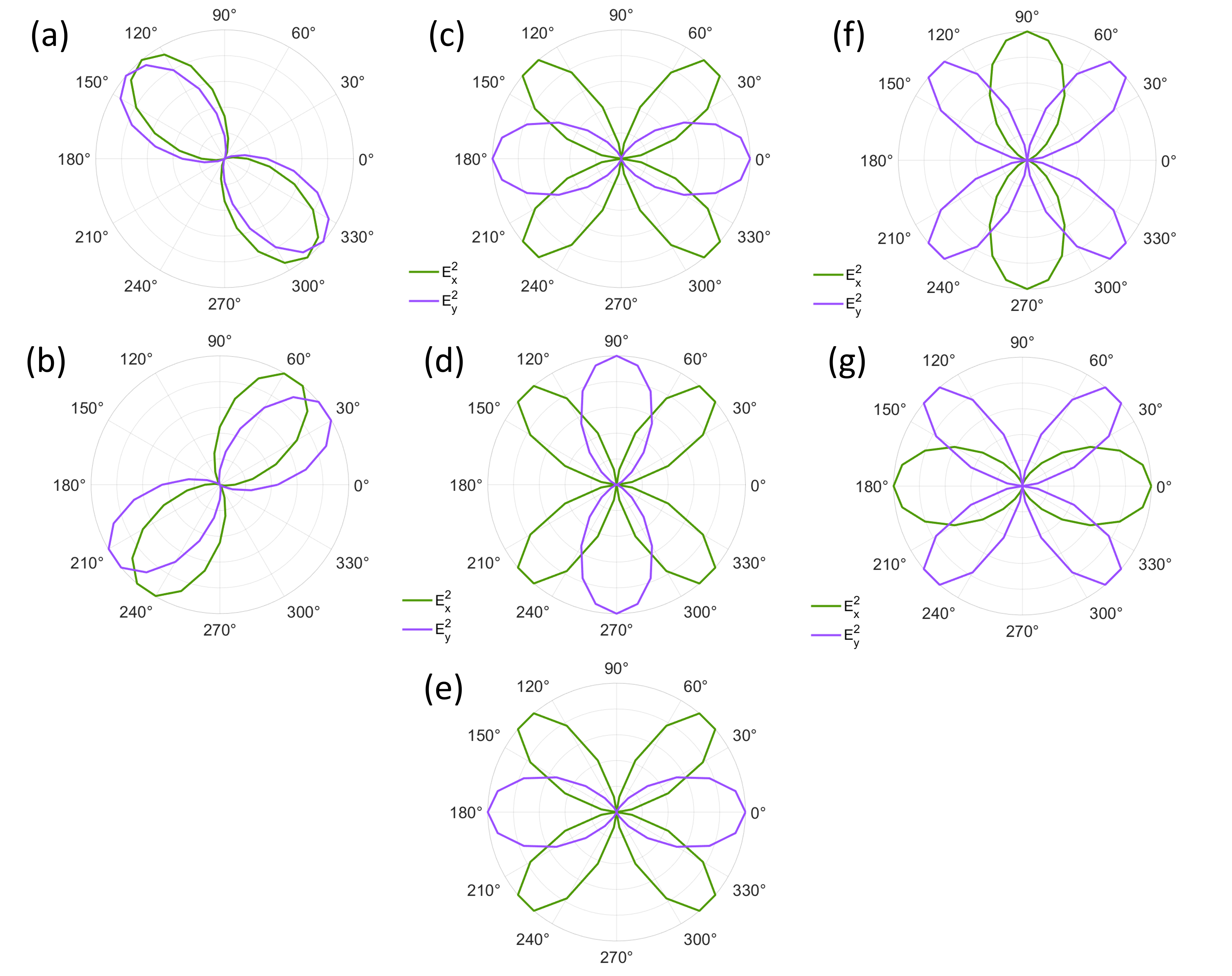}
\caption{The $\tau=150$ fs THz pulse as it appears in the Fourier-transformed squared $x$- or $y$-component of the electric field, with elapsed time of the simulation $t_{final}=500$ fs, as a function of the incident electric field polarization. As in Fig.~\ref{fig:Figure4}, the green line represents the $x$-component and the purple line represents the $y$-component. Both components in each plot are normalized for easier comparison. (a,c,f) The incident electric field is pumped at the $v$ resonance of each geometry, with (a) both components being equal in magnitude for the LS, (c) the $y$-component being two orders of magnitude stronger for the SRR, and (f) the $x$-component being two orders of magnitude stronger for the NC. (b,d,g) The incident field is pumped at $u$ resonance of each geometry, with (b) both components being equal in magnitude, (d) the $y$-component being one order of magnitude stronger, and (g) the $x$-component being more than two orders of magnitude stronger. (e) The system is pumped at the $q$ resonance of the SRR, with the $y$-component being two orders of magnitude stronger.}
\label{fig:Figure5}
\end{figure*}

Fig.~\ref{fig:Figure5} examines signal calculated at the maximum of the THz peak for each geometry as a function of the incident pulse polarization angle $\theta$. Interestingly, we see the stronger electric field component corresponding to the $u$ resonance of each geometry, that is, no matter the resonant angle of polarization, the dominant $E^2$ component is the same. For example, $E_y^2$ is the dominant component for all three resonances of the SRR, regardless of the polarization angle for each resonance. This is noticeably prevalent in the SRR and NC geometries, as their resonant polarization angles are perpendicular to one component (i.e. $E_y$ for the SRR and $E_x$ for the NC) while being parallel to the other. One could hypothesize this is the same for the LS geometry, except the difficulty comes from its resonant polarization angle, which is equidistant from both $x$- and $y$-components, being $\theta=135^{\circ}$, or $45^{\circ}$ from both the $x$- and $y$-axes. One can envision a further exploration of the dominant component through geometries that with resonant polarization angles that favor a particular component without being exactly perpendicular to one component (and subsequently parallel to the other). 

\section{Analysis of transmitted pulse polarization and lineshape}
\label{Sec:TheoryLinSh}

To gain insights into the results of the numerical simulations discussed in the preceding sections, here, we discuss a theoretical model for both SHG and THz pulse generation, considering the array of metal nanoparticles as an effective plasmonic medium characterized by the linear refractive index and second-order nonlinear susceptibility. Below, we explore how the symmetries of the nonlinear susceptibility tensor define the relationship between the incident pump polarization and the transmitted signal polarization. We also discuss the lineshape and temporal profile of the generated THz pulse.

\subsection{Analytical model}
\label{Sec:TheoryModel}

All the expression below are derived in the natural $(u,v)$-coordinate system aligned with the surface plasmon eigenmode polarization. In this frame of reference, the linear response of the layer is described by complex refractive index $n(\omega)=(n_u(\omega),n_v(\omega))$, possessing anisotropy associated with the orthogonal orientation of the two local surface plasmon modes (compare panels b, c, e, f, and h, i of Fig.~\ref{fig:Figure1}). The nonlinear response accounting for the same surface plasmon resonances is described by the second order susceptibility tensor $\chi_{msl}^{(2)}(\omega;\omega_1,\omega_2)$ with the indices $m,s,l\in \{u,v\}$.

Next, we introduce an optical pump incident perpendicularly onto the surface of the metal nanoparticle (MNP) layer of thickness $L_z$ at $z=L_z/2$, characterized by a wave vector $k_m(\omega) = n_m(\omega) \omega^2/c^2$ aligned along the negative $z$-direction. The pump electric field is linearly polarized, and represented as a vector $\bm{E}_\text{pump}(z, \omega) = ({\cal E}^{(p)}_u(z, \omega)e^{-i k_u(\omega) z}, {\cal E}^{(p)}_v(z, \omega)e^{-i k_v(\omega) z}, 0)$. Here, ${\cal E}^{(p)}_m(z, \omega)$ is the pulse envelope function defined inside the array, $-L_z/2 \leq z\leq L_z/2$. For the processes of the SHG and DFG, the induced nonlinear polarization of the MNP layer can be described as the following spectral convolution of the pump pulse and the second-order susceptibility
\begin{eqnarray}
\label{PNL-def}
P^{(2)}_m(z,\omega) &=& \varepsilon_o\iint d\omega_1 d\omega_2\delta(\omega-\omega_1\mp \omega_2)
\\\nonumber &\times&
   \chi_{msl}^{(2)}(\omega;\omega_1,\pm \omega_2)        
    {\cal E}^{(p)}_s(z,\omega_1)
\\\nonumber &\times&
    {\cal E}^{(p)}_l(z,\pm\omega_2)
    e^{-i[k_s(\omega_1)\pm k_l(\pm\omega_2)]z}.
\end{eqnarray}
In Eq.\eqref{PNL-def}, the term $\chi_{msl}^{(2)}(\omega;\omega_1,\omega_2)$ stands for the SHG, while $\chi_{msl}^{(2)}(\omega;\omega_1,-\omega_2)$ describes DFG resulting in the THz signal. The discussion below assumes summation over the repeated indices $s$ and $l$ running through the $(u,v)$-coordinates. The delta-function under the integral enforces the conservation of photon energy during the SHG and DFG process. By integrating over $d\omega_1$, we simplify Eq.(\ref{PNL-def}) to the form
\begin{eqnarray}
\label{PNL-simp}
&~&\hspace{-1cm} P^{(2)}_m(z,\omega) = \varepsilon_o\int d\omega' \chi_{msl}^{(2)}(\omega;\omega\mp\omega',\pm\omega')
\\\nonumber &~&\times            
    {\cal E}^{(p)}_s(z,\omega\mp\omega'){\cal E}^{(p)}_l(z,\pm\omega')
    e^{-i[k_s(\omega+\omega')\pm k_l(\pm\omega')]z}.
\end{eqnarray}

The signal electric field $E_m(z,\omega)={\cal E}_m(z,\omega)e^{-ik(\omega)z}$ generated by the pump inside the plasmonic medium satisfies the Helmholtz equation 
\begin{eqnarray}
\label{Heltz-def}
\left[\partial^2_z+k^2(\omega)\right]{\cal E}_m(z,\omega)e^{-ik_m(\omega)z}=-\frac{\omega^2}{c^2\varepsilon_o}P^{(2)}_m(z,\omega). 
\end{eqnarray}
Upon substituting  Eq.~(\ref{PNL-simp}) into the right-hand side of Eq.~(\ref{Heltz-def}) and applying the slowly varying amplitude approximation, the equation transforms into
\begin{eqnarray}
\label{Heltz-chi2}
&~&\hspace{-1.5cm}\partial_z {\cal E}_m(z,\omega)=\frac{i\omega}{2n(\omega)c}
    \int d\omega' \chi_{msl}^{(2)}(\omega;\omega\mp\omega',\pm\omega')
\\\nonumber &~&\times         
     {\cal E}^{(p)}_s(z,\omega\mp\omega'){\cal E}^{(p)}_l(z,\pm\omega')
    e^{i\Delta k_{msl}(\omega,\omega')z}. 
\end{eqnarray}
Here, we use the following shorthand notation for the wavevector mismatch
\begin{eqnarray}
\label{dk-def}
\Delta k_{msl}(\omega,\pm\omega') \equiv k_m(\omega)-k_s(\omega\mp \omega')\mp k_l(\pm\omega'), 
\end{eqnarray}
where the first wavevector $k_m(\omega)$ is a real quantity, assuming that the refractive index has no resonances in the second harmonic UV and THz regions. 

Assuming that a strong pump field propagating through a thin layer of MNP remains unchanged, Eq.~(\ref{Heltz-chi2}) can be integrated over $dz$. Given the boundary condition ${\cal E}_m(L_z/2,\omega)=0$, indicating no incident THz field, the resulting transmitted THz field is 
\begin{eqnarray}
\label{Et-sinc}
&~&\hspace{-0.5cm}{\cal E}^{(t)}_m(\omega)=\frac{i\omega}{n_m(\omega)c}\int d\omega' 
    \tilde\chi_{msl}^{(2)}(\omega;\omega\mp\omega',\pm\omega')
\\\nonumber &~&\times
        {\cal E}^{(i)}_s(\omega\mp \omega'){\cal E}^{(i)}_l(\pm\omega')
    \frac{\sin\left(\Delta k_{msl}(\omega,\pm\omega')L_z/2\right)}{\Delta k_{msl}(\omega,\pm\omega')}. 
\end{eqnarray}
In Eq.~\eqref{Et-sinc}, we introduced effective second order susceptibility 
\begin{eqnarray}
\label{chi2-eff}
&~&\hspace{-0.5cm}\tilde\chi_{msl}^{(2)}(\omega;\omega\mp\omega',\pm\omega')\equiv
\\\nonumber &~&t_m^{(t)}(\omega)\chi_{msl}^{(2)}(\omega;\omega\mp\omega',\pm\omega')
    t_{s}^{(i)}(\omega \mp\omega')t_{l}^{(i)}(\pm\omega'),
\end{eqnarray}
which contains the transfer matrix elements 
\begin{eqnarray}
\label{tt-def}
t_m^{(t)}(\omega) &=& \frac{2 n_m(\omega)}{n_m(\omega)+1}e^{-i[k_m(\omega)-\omega/c]L_z},
\\\label{ti-def}
t_m^{(i)}(\omega) &=& \frac{2 [n_m(\omega)+1]}{[n_m(\omega)+1]^2-[n_m(\omega)-1]^2 e^{-2ik_m(\omega) L_z }},
\end{eqnarray}
accounting for the boundary conditions at $z=L_z/2$ and $z=-L_z/2$, respectively.  To obtain this expression, we replaced the pump field, ${\cal E}^{(p)}_m(\omega)$, inside the plasmonic medium with the incident pump field, ${\cal E}^{(i)}_s(\omega)$, outside plasmonic medium, $z>L_z/2$, according to the linear relationship ${\cal E}^{(p)}_s(\omega)=t_s^{(i)}(\omega){\cal E}^{(i)}_s(\omega)$. Similarly, the transmitted field, ${\cal E}^{(t)}_m(\omega)$, outside the medium, $-z<L_z/2$ has been introduced via the relationship ${\cal E}^{(t)}_m(\omega)=t_m^{(t)}(\omega){\cal E}_m(\omega)$, where ${\cal E}_m(\omega)$ is the generated field at $z=L_z/2$ calculated by integrating Eq.~\eqref{Heltz-chi2}. 

The simulations of SHG discussed in Sec.~\ref{Sec:LinearSHG} were performed in the CW regime. This regime can be represented in Eq.\eqref{Et-sinc} by setting the incident electric field envelopes to ${\cal E}^{(i)}_s(\omega) = {\cal A}_s \delta(\omega - \omega_0)$, where ${\cal A}_s$ is the real amplitude of the incident electric field and $\omega_0$ is the central frequency within the optical range. By subsequently integrating Eq.~\eqref{Et-sinc} over the frequency, we obtain the well-known expression for the transmitted electric field due to SHG
\begin{eqnarray}
\label{SHG-CW}
\hspace{-0.5cm}{\cal E}^{(t)}_m(2\omega_0)&=&\frac{2\omega_0i}{n_m(2\omega_0)c} 
    \tilde\chi_{msl}^{(2)}(2\omega_0;\omega_0,\omega_0) {\cal A}^{(i)}_s{\cal A}^{(i)}_l
\\\nonumber &~&\times
     \frac{\sin\left(\Delta k_{msl}(2\omega_0,\omega_0)L_z/2\right)}{\Delta k_{msl}(2\omega_0,\omega_0)}. 
\end{eqnarray}

From now on and for the rest of this subsection, we will consider the transmitted THz field by making the appropriate sign choice in Eq.~\eqref{Et-sinc} and remembering that ${\cal E}^{(p)}_l(z,-\omega_2) = {\cal E}^{(p)}_l(z,\omega_2)$ and $k_l(-\omega_2) = k^*_l(\omega_2)$. This leads to the expression
\begin{eqnarray}
\label{ETHz-sinc}
&~&\hspace{-0.5cm}{\cal E}^{(t)}_m(\omega)=\frac{i\omega}{n_m(\omega)c}\int d\omega' 
    \tilde\chi_{msl}^{(2)}(\omega;\omega+\omega',-\omega')
\\\nonumber &~&\times
        {\cal E}^{(i)}_s(\omega + \omega'){\cal E}^{(i)*}_l(\omega')
    \frac{\sin\left(\Delta k_{msl}(\omega,-\omega')L_z/2\right)}{\Delta k_{msl}(\omega,-\omega')}. 
\end{eqnarray}
where the incident pump frequency distribution is modelled using the Gaussian function
\begin{eqnarray}
\label{Ep-Gauss}
{\cal E}^{(i)}_s(\omega) = \frac{{\cal A}_s}{\sqrt{\pi^{1/2}\sigma_\text{p}}}\exp\left(-\frac{[\omega-\omega_0]^2}{2\sigma_\text{p}^2}\right). 
\end{eqnarray}
In Eq.~\eqref{Ep-Gauss}, ${\cal A}_s$ is a real amplitude of the incident electric field, $\omega_0$ is the central frequency, and $\sigma_\text{p}$ is the spectral width of the pulse. It is noteworthy that in Eq.~\eqref{ETHz-sinc} the last argument in the susceptibility tensor has negative sign, i.e., $-\omega_0$. This indicates the fact that the generated signal frequency is determined by the difference in the frequencies of the incident pump field components, namely, $ {\cal E}^{(i)}_s(\omega + \omega')$ and ${\cal E}^{(i)*}_l(\omega')$. Provided the incident pump is CW, our model recovers the well known limit of optical ratification.

Equations~(\ref{ETHz-sinc}) and \eqref{chi2-eff}, supplemented by Eq.~(\ref{Ep-Gauss}), can be numerically integrated given a model for the linear $n(\omega)$ and nonlinear $\chi^{(2)}$ responses of the MNP array. The frequency dependence (dispersion) of these parameters should influence the lineshape of the output pulse. To examine the effect of dispersion on the THz pulse line shape, we approximately calculate the integral on the right-hand side of Eq.~(\ref{ETHz-sinc}). We assume that the frequency of the pump pulse is nearly resonant with the surface plasmon resonance ($\omega_0 \sim \omega_m$, where $m\in\{u,v\}$) and the spectral width of the pump pulse is much narrower than the width of the plasmon resonance ($\sigma_\text{p}\ll\gamma_m$, where $m\in\{u,v\}$). In this case, we account for the frequency dependence of the wave vector and second-order susceptibility by expanding them into the Taylor series near the pump central frequency $\omega_0$ up to the second order in the frequency deviation $\delta\omega=\omega_0-\omega'$. After performing the calculation outlined in Appendix~\ref{Appx:SaddlePointInt}, we arrive at the approximate representation of the generated THz field
\begin{eqnarray}
\label{ETHz-approx}
{\cal E}^{(t)}_m(\omega)&=&\frac{L_z\omega~e^{-g_{msl}(\omega)}}{2cn_m(\omega)
    \sqrt{1+\frac{\sigma_\text{p}}{\sigma_{msl}(\omega)}}}
\\\nonumber&\times&
    \tilde\chi_{msl}^{(2)}(\omega;\omega+\omega_0,-\omega_0)
   {\cal A}_s{\cal A}_l, 
\end{eqnarray}
containing the THz pules lineshape function which reads
\begin{eqnarray}
\label{gnsl-def}
&~&\hspace{-1cm} g_{msl}(\omega)=\kappa_{msl}(\omega)+\left[
 \frac{\omega^2}{4\sigma^2_\text{p}}\left(1+\frac{2\sigma^2_\text{p}}{\sigma^2_{msl}(\omega)}\right)
\right.\\\nonumber&-&\left.
    \frac{\omega\tau_{msl}(\omega)}{2}
    -\frac{\sigma_\text{p}^2\tau_{msl}^2(\omega)}{4}\right]
 \left[1+\frac{\sigma^2_\text{p}}{\sigma^2_{msl}(\omega)}\right]^{-1}. 
\end{eqnarray}
The lineshape function contains $\kappa_{msl}(\omega)$, $\tau_{msl}(\omega)$, and $\sigma^{-2}_{msl}(\omega)$ whose explicit representation in terms of the wavevector mismatch and second-order susceptibility are given in Appendix~\ref{Appx:SaddlePointInt} by Eqs.~\eqref{kappa-msl-def}, \eqref{tau-msl-def}, and \eqref{sigma-msl-def}, respectively.

Moreover, the leading contribution to the THz pulse lineshape can be obtained by neglecting the dispersion effects in the vicinity of the pump pulse frequency. This means setting $\kappa_{msl}=\tau_{msl}=\sigma^{-2}_{msl}=0$ in Eqs.(\ref{ETHz-approx})- and (\ref{gnsl-def}) to obtain
\begin{eqnarray}
\label{ETHz-0}
{\cal E}^{(t)}_m(\omega)&=&\frac{L_z\omega~e^{-\frac{\omega^2}{4\sigma^2_\text{p}}}}{2cn_m(\omega)}
\tilde\chi_{msl}^{(2)}(\omega;\omega+\omega_0,-\omega_0) {\cal A}_s{\cal A}_l. 
\end{eqnarray}
Further simplification of this expression can be performed by making the same as in Appendix~\ref{Appx:SaddlePointInt} second order expansion of the refractive index and the nonlinear susceptibility in the THz region, i.e., near $\omega=0$. This calculation results in
\begin{eqnarray}
\nonumber
&~&\hspace{-1cm}{\cal E}^{(t)}_m(\omega)=\frac{L_z\omega}{2cn_m(0)}
\exp\left( -\frac{\omega^2}{4\sigma^2_\text{p}}\left[1-\frac{\sigma^2_\text{p}}{\sigma^{(0)2}_{nsl}}\right]
    +\tau^{(0)}_{msl}\omega \right)
\\\label{ETHz-0-Gauss}&~&~~\times    
    \tilde\chi_{msl}^{(2)}(0;\omega_0,-\omega_0){\cal A}_s{\cal A}_l, 
\end{eqnarray}
where
\begin{eqnarray}
\label{tau0-def}
&~&\hspace{-1cm}\tau^{(0)}_{msl}= \left[-\frac{\partial_\omega n_m(\omega)}{n_m(\omega)} 
    +\frac{\partial_\omega \tilde\chi_{msl}^{(2)}(\omega;\omega+\omega_0,-\omega_0)}
        { \tilde\chi_{msl}^{(2)}(\omega;\omega+\omega_0,-\omega_0)}\right]_{\omega=0},
\\\label{sigma0-def}
&~&\hspace{-1cm}\frac{1}{\sigma^{(0)2}_{nsl}}= -2\tau^{(0)2}_{msl}
\\\nonumber &~&
-2\left[\frac{\partial^2_\omega n_m(\omega)}{n_m(\omega)} 
    -\frac{\partial^2_\omega \tilde\chi_{msl}^{(2)}(\omega;\omega+\omega_0,-\omega_0)}
        {\tilde\chi_{msl}^{(2)}(\omega;\omega+\omega_0,-\omega_0)}\right]_{\omega=0}.
\end{eqnarray}
It is noteworthy that a more general expression of the transmitted field given by Eq.~\eqref{ETHz-approx} can be brought to a similar functional form by performing the expansion of the lineshape function~\eqref{gnsl-def} and the square root normalization factor up to the second order in $\omega$. The result will differ from that above by much more complex dependence of the parameters $\tau^{(0)}_{msl}$ and $\sigma^{(0)-2}_{msl}$ on the wave vector mismatch and second order susceptibility compared to Eqs.~\eqref{tau0-def} and \eqref{sigma0-def}. Since this gains no additional insights into the lineshape function structure, we do not provide this explicit representation but rather examine the deviations in the pulse spectrum numerically.

Finally, Eq.~(\ref{ETHz-0-Gauss}) can be Fourier transformed to the time domain to obtain the following temporal representation for the transmitted THz pulse
\begin{eqnarray}
\nonumber
&~&\hspace{-1cm}{\cal E}^{(t)}_m(t)=\frac{\sqrt{2}L_z\sigma^4_\text{p}\left[\tau^{(0)}_{msl}-it\right]}{cn_m(0)\left[1-\frac{\sigma^2_\text{p}}{\sigma^{(0)2}_{msl}}\right]^{3/2}}
\exp\left(-\frac{\sigma^2_\text{p}\left[t+i\tau^{(0)}_{msl}\right]^2}{1-\sigma^2_\text{p}/\sigma^{(0)2}_{nsl}}
 \right)
 \\\label{ETHz-0-time}&~&
   \times\tilde\chi_{msl}^{(2)}(0;\omega_0,-\omega_0){\cal A}_s{\cal A}_l. 
\end{eqnarray}

\subsection{Analysis of polarization dependence}
\label{Sec:PolDep}

Let us introduce the refractive index 
\begin{equation}
\label{nw-chi1}
n_m(\omega)=\sqrt{1+\chi_{m}^{(1)}(\omega)},
\end{equation}
where the linear susceptibility is modeled using the Lorentzian lineshape
\begin{equation}
\label{eq:lorentzian}
    \chi^{(1)}_{m}(\omega) = \frac{f_m}{\hbar^2\left(\omega_{m}^2 - \omega^2 - i\omega\gamma_m\right)}.
\end{equation}
Here, $f_m$, $\omega_m$, and $\gamma_m$, represent the oscillator strength, resonant frequency, and associated dephasing rate of the surface plasmon mode with polarization $m\in\{u,v\}$, respectively. These parameters were determined for the LS MNP geometry by fitting~\cite{Shah:24} associated curves $u$ and $v$ of Fig.~\ref{fig:Figure1}b using Eq.~\eqref{eq:lorentzian}. The results are provided in Table~\ref{table:linear_fit}.

\begin{table}[t]
\centering
\begin{tabular}{ |c|c|c|c| } 
\hline
 $m$ & $f \ (\text{eV}^2)$ & $\hbar \omega_m$ (eV)   & $\hbar\gamma_m$ (eV) \\
\hline
$v$ & 1.194& 1.257 & 0.1445 \\ 
$u$ & 1.27 & 1.984& 0.145 \\  
\hline
\end{tabular}
\caption{The oscillator strength $f_m$, resonant frequency $\omega_m$, and dephasing parameter $\gamma_m$ obtained for the LS MNP by fitting curve $v$ and $u$ of Fig.~\ref{fig:Figure1}~b using the Lorentzian model for the linear susceptibility given in Eq.~\eqref{eq:lorentzian}.}
\label{table:linear_fit}
\end{table}

Following our previous work on LS MNPs~\cite{Shah:24}, we approximate the second-order susceptibility as the product of the linear susceptibilities defined in Eq.~\eqref{eq:lorentzian}, weighted by the anharmonicity tensor, $A_{msl}$,
\begin{equation}
\label{chi-A}
    \chi^{(2)}_{msl}(\omega_3, \omega_2, \omega_1) = \chi^{(1)}_m(\omega_3)A_{msl} \chi^{(1)}_s(\omega_2) \chi^{(1)}_l(\omega_1).
\end{equation}
Under the condition of C$_{2v}$ point group symmetry, this tensor has four nonzero (and three independent) elements $\{A_{uuu}, A_{uvv}, A_{vvu}=A_{vuv}\}$.\footnote{In the absence of dissipation $A_{uvv} = A_{vvu}$.} By fitting the SHG data for the LS MNPs presented in Fig.\ref{fig:Figure4}~b and c using Eq.~\eqref{SHG-CW}, we extracted the off-diagonal components of this tensor, normalized to $A_{uuu}$. The result is $A_{vvu}/A_{uuu} = 1$, and $A_{uvv}/A_{uuu} = 2$. The value of the diagonal element $A_{uuu}$ is not critical for further analysis and was not extracted.

\begin{figure}[t]
\centering
\includegraphics[width=0.3\textwidth]{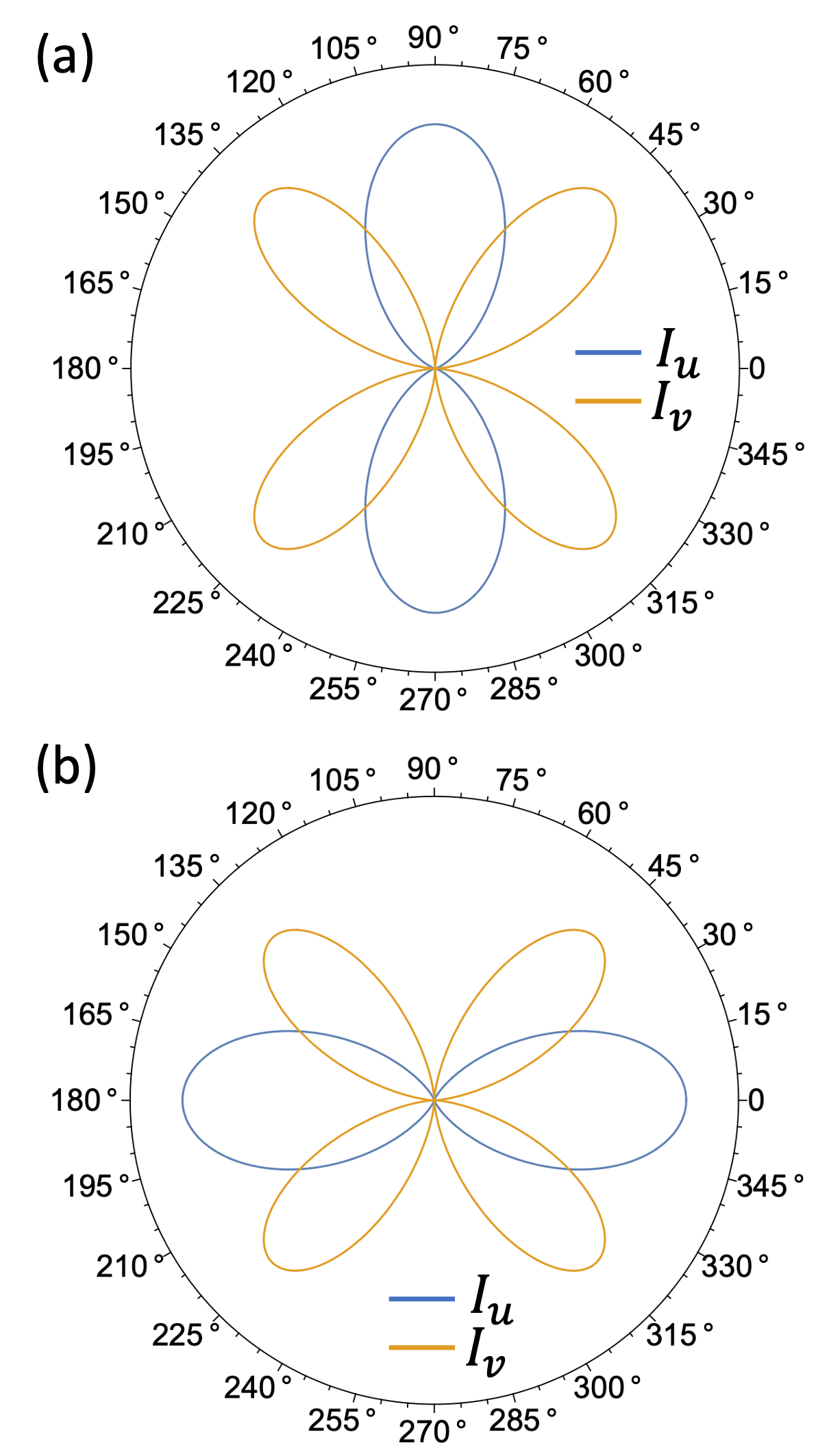} 
\caption{Normalized SHG intensity along natural $(u,v)$-axis as function of pump polarization for LS. (a) Pump set to $\omega_0\omega_v$ and (b) $\omega_0=\omega_u$.}
\label{fig:Figure7}
\end{figure}

Now, we turn our attention to the polarization dependence of the SHG signal presented in Fig.~\ref{fig:Figure4}. Considering the non-zero components of the second-order susceptibility tensor, we identify their contributions to the transmitted electric field through the following contractions with the incident electric field components
\begin{align}
 \label{PuvsChi}
    {\cal E}^{(t)}_u&\sim \tilde\chi_{uuu}^{(2)}{\cal E}_u^{(i)}{\cal E}^{(i)}_u
    +\tilde\chi_{uvv}^{(2)}{\cal E}_v^{(i)}{\cal E}^{(i)}_v,
\\\label{PvvsChi} 
    {\cal E}^{(t)}_v&\sim \tilde\chi_{vvu}^{(2)}{\cal E}_v^{(i)}{\cal E}^{(i)}_u
    +\tilde\chi_{vuv}^{(2)}{\cal E}_u^{(i)}{\cal E}^{(i)}_v.
\end{align}
Moreover with the help of Eq.~\eqref{SHG-CW}, we calculated their intensities $I_u\sim|{\cal E}^{(t)}_u|^2$ and $I_v\sim|{\cal E}^{(t)}_v|^2$ as a function of the incident pump field polarization angle defined with respect to the $u$-axis. The results of the normalized transmitted signal for the pump central frequency tuned in resonance with the $u$ and $v$ surface plasmon modes are shown in panels~a and b of Fig.~\ref{fig:Figure7}, respectively. 

Taking into account that the natural frame of reference is $\pi/4$ rotated with respect to the laboratory one (Fig.~\eqref{fig:Figure1}~a), the fields in the latter one are linear superposition of the fields in Eq.~\eqref{PuvsChi} and ~\eqref{PvvsChi}
\begin{align}
\label{Ex-LS}
    {\cal E}^{(t)}_x &= \frac{1}{\sqrt{2}} ({\cal E}^{(t)}_u + {\cal E}^{(t)}_v)
\\\label{Ey-LS}
    {\cal E}^{(t)}_y &= \frac{1}{\sqrt{2}} (-{\cal E}^{(t)}_u + {\cal E}^{(t)}_v).
\end{align}
Accordingly, the interference of these fields transforms the angular distributions shown in Fig.~\ref{fig:Figure7} to the form presented Fig.~\ref{fig:Figure4}~a, b. 

Interestingly, symmetry arguments also enable us to explain the normalized angular diagrams for the SRR and NC geometries shown in panels c, d, f, and g of Fig.\ref{fig:Figure4} using the LS MNP parameterization. However, we do not address the quadruple resonance of the SRR MNPs presented in Fig.\ref{fig:Figure4}~e, as our model does not account for this type of response. According to Fig.\ref{fig:Figure1}~g, the natural coordinate system for the NC MNP coincides with the laboratory coordinate system. Consequently, the angular distributions shown in Fig.\ref{fig:Figure7}~a, b match those in Fig.\ref{fig:Figure4}~f, g, resepcetively. For the SSR geometry, the natural coordinate system is rotated by $\pi/2$ with respect to the laboratory coordinates (Fig.\ref{fig:Figure1}~d). Thus, identifying the fields in the laboratory coordinate system as ${\cal E}^{(t)}_x=-{\cal E}^{(t)}_v$ and ${\cal E}^{(t)}_y={\cal E}^{(t)}_u$. This explains the correspondence between panel~a (panel~b) of Fig.~\ref{fig:Figure7} and panel~d (panel~c) of Fig.~\ref{fig:Figure4}. The main point of this analysis is that, for all the geometries considered, the angular distributions of the transmitted SHG signal calculated in the natural frame of reference (Fig.~\ref{fig:Figure7}) are the same. The variations observed in Fig.~\ref{fig:Figure4} are due to different orientations of the laboratory frame of reference. This fact is important to consider when explaining experimental data.

Our analysis of the angular distributions for the SRR and NC geometries is based on the model parameterized for the LS MNP. This emphasizes that the shape of the diagrams is primarily dominated by the C$_{2v}$ symmetry, which identifies the nonzero contributions of the susceptibility tensor as presented in Eqs.~\eqref{PuvsChi} and \eqref{PvvsChi}. The frequency dependence of the tensor and the phase matching conditions play a minor role due to the similarity of the linear response associated with each geometry, as seen by comparing panels b, e, and h of Fig.\ref{fig:Figure1}.

\begin{figure}[t]
\centering
\includegraphics[width=0.5\textwidth]{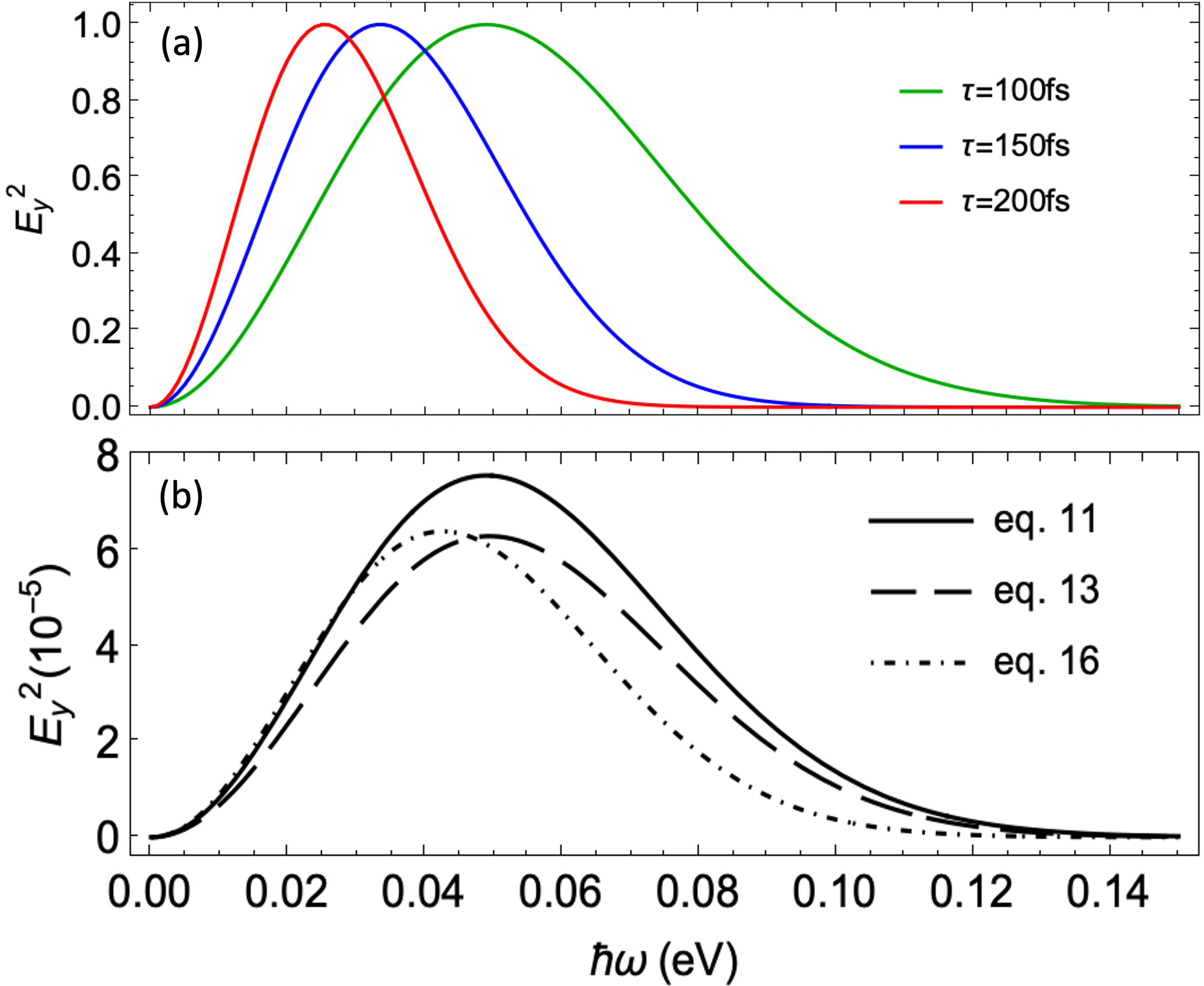} 
\caption{(a) Theoretical modeling for the line-shape of the THz pulse for LS MNP using Eq.~\ref{ETHz-sinc}. Three curves represent normalized $y$-component of the THz intensity at different values of the pump pulse duration $\tau$. (b) Comparison of approximate models given by  Eq.~\eqref{ETHz-approx} and Eq.~\eqref{ETHz-0-Gauss}, with the exact model given by Eq.~\eqref{ETHz-sinc} for the pump duration set to $\tau=100~fs$.}
\label{fig:Figure8}
\end{figure}

The angular distributions for the THz signal have been calculated using Eq.\eqref{ETHz-sinc}, and obtained normalized intensities match those presented in Fig.\eqref{fig:Figure7}. This explains the close resemblance between the distributions observed in Figs.~\ref{fig:Figure4} and \ref{fig:Figure5} for the SHG and THz signals, respectively. Consequently, we can draw similar conclusions regarding the angular distribution of the THz field as those provided earlier for the SHG. However, to reproduce the offset between the purple and green curves observed in Figs.\ref{fig:Figure5}~a and b, we had to adjust the anharmonicity tensor component ratio from $A_{uvv}/A_{uuu} = 2$ to $A_{uvv}/A_{uuu} = 1$ leaving the $A_{vvu}/A_{uuu} = 1$ intact. This change affects the relative values of the electric field amplitudes that interfere according to Eqs.\ref{Ex-LS} and \ref{Ex-LS}, resulting in the angular dependence observed in Figs.\ref{fig:Figure5}~a and b. The analysis indicates that the anharmonicity tensor is frequency-dependent, as captured by the hydrodynamic model. We account for this dependence in the following calculation of the THz lineshape by adopting the new aforementioned set of values for the anharmonicity tensor components.


\subsection{Analysis of THz lineshape}

\begin{figure}[t]
\centering
\includegraphics[width=0.4\textwidth]{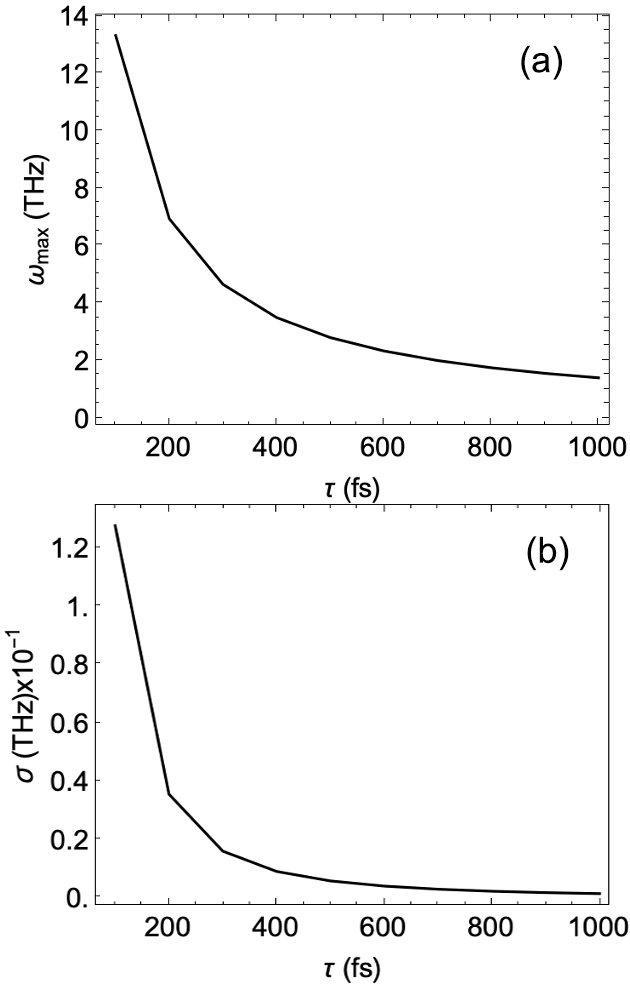} 
\caption{First moment of the THz lineshape as a function of pump pulse duration.}
\label{fig:Figure9}
\end{figure}


Below, we use our model developed in Sec.\ref{Sec:TheoryModel} and parameterized in Sec.\ref{Sec:PolDep} to examine the lineshape of the THz signal due to the LS MNP. Specifically, we focus on the case shown in Fig.\ref{fig:Figure5a}~c, where the pump pulse is tuned to resonate with the $v$-mode, and the detected field intensity is projected onto the $y$-axis of the laboratory coordinate system. According to Eqs.~\eqref{PuvsChi},  \eqref{PvvsChi}, and \eqref{Ey-LS}, the only non-vanishing contribution to this signal comes from the $\tilde\chi_{uvv}^{(2)}$ tensor component. Therefore, this tensor component is used in all the following calculations.

Figure~\ref{fig:Figure8}~a presents the transmitted THz pulse spectrum calculated according to Eq.~\eqref{ETHz-sinc}, using a Fourier transform-limited Gaussian pulse with a time-bandwidth product of 0.882. This calculation accurately reproduces the results obtained in Fig.~\ref{fig:Figure5a}~c, further validating our analytical model.
Additionally, in panel b, we compare the approximate results for the THz spectra determined by Eqs.~\eqref{ETHz-approx} and \eqref{ETHz-0-Gauss} with the exact one (Eq.~\eqref{ETHz-sinc}) for a pump pulse duration of 100 fs. The plot shows that accounting for the dispersion effect of the pump pulse near the surface plasmon resonance (Eq.~\eqref{ETHz-approx}) accurately reproduces the peak frequency of the THz pulse and provides an adequate approximation for the spectral widths. Neglecting this effect (Eq.~\eqref{ETHz-0-Gauss}) introduces more discrepancy but still offers a fair approximation. Our calculations (not shown in the plot) demonstrate that the discrepancy between the models' predictions diminishes as the pump pulse duration increases.

\begin{figure}[t]
\centering
\includegraphics[width=0.4\textwidth]{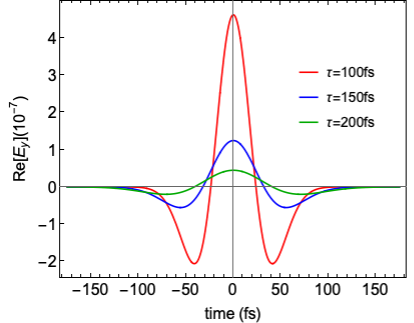} 
\caption{Temporal profile of the real part of transmitted THz pulse calculated for the  LS MNP using Eq.~\eqref{ETHz-0-time}. The three curves represent different excitation pulse duration corresponding to the spectra shown in Fig.~\ref{fig:Figure8}~(a).}
\label{fig:Figure10}
\end{figure}

Having validated the approximations, we further examine how the spectrum maximum and width depend on the pump pulse duration by calculating the first and second moments of the spectrum defined in Eq.~\eqref{ETHz-approx}. The results are presented in Fig.\ref{fig:Figure9}. This plot shows that as the pump duration increases, the spectral maximum shifts towards zero frequency, and the linewidth narrows. This behavior can be explained by the fact that the THz signal results from the DFG between the pulse spectral components. Specifically, as the pulse duration increases, its spectral width decreases, leading to a narrower THz signal. Additionally, this trend confirms that in the CW limit, characterized by long pulse duration, the optical rectification limit characterized by static polarization should be recovered. To complete our analysis, we also calculated the temporal profile of the transmitted THz pulse using Eq.~\eqref{ETHz-0-time}, and the results are shown in Fig.~\ref{fig:Figure10}. The trend of temporal pulse narrowing as the pump duration increases is clearly evident in this plot.

\section{Conclusion}\label{Sec:conclusion}

We presented the systematic studies of the optical characteristics of plasmonic nanoparticles obeying C$_{2v}$ symmetry. Our main results yield significant insights into both particles' linear and nonlinear optical properties. By examining several geometric configurations, it was found that each configuration displays two prominent plasmon bands aligned with the axes of symmetry and a series of higher-frequency resonances associated with quadrupole modes and features related to their sharp corners.

The advanced semiclassical hydrodynamic model was employed to gain insights into the nonlinear optical behavior of these nanoparticles, specifically focusing on the second-order angular resolved optical response. 
Under CW pumping, our simulations revealed that SHG is highly dependent on the nanoparticle's geometry and the polarization of the incident wave. Evidently, the inherent C$_{2v}$ symmetry of the nanoparticles plays a crucial role in defining the polarization states and selection rules for the SHG signal, highlighting the intricate interaction between nanoparticle's symmetry and its nonlinear optical responses.

When exploring pulsed excitations, the phenomenon of broadband THz generation was observed. The THz emission spectra showcased distinctive features linked to the plasmonic resonances and symmetry of the nanoparticles, with the polarization patterns of the emitted THz waves providing further evidence of the impact of nanoparticle geometry on the second-order optical processes.

The use of the analytical theory, which aligns closely with the results from numerical experiments, elucidates the physical mechanisms behind THz radiation generation in these systems. Our theory illustrates how the mixing of various frequency components of the incident pulse via the second-order nonlinear susceptibility leads to THz emission, and an expression for the far-field THz intensity was derived, connecting it to the incident pulse parameters and the nonlinear response tensor of the nanoparticle.

Our findings demonstrate the potential of these structures in applications such as nonlinear spectroscopy, nanophotonics, and optoelectronics. The strong SHG response and efficient THz generation observed suggest these nanoparticles are promising candidates for future technological applications, while the theoretical framework developed provides a robust basis for predicting the behavior of similar nanostructures in nonlinear optical contexts.

\section{Acknowledgements}
The authors are grateful to Prof. Joseph Zyss for fruitful discussions. This work by M.R.C and M.S. is supported by the Air Force Office of Scientific Research under Grant No. FA9550-22-1-0175. Theoretical and computational work is performed, in
part, at the Center for Integrated Nanotechnologies, an Office of Science User Facility operated for the
U.S. Department of Energy (DOE) Office of Science by Los Alamos National Laboratory (Contract 89 233 218CNA000001) and Sandia National Laboratories (Contract DE-NA-0003525) (user project No. 2021BC0087). Work by S.S. is supported  by the Center of Nonlinear Studies via Laboratory Directed Research and Development program of Los Alamos National Laboratory under project number 3W400A-XXJ4. A.P. (M.R.C.) research was supported (in part) by the Laboratory Directed Research and Development program of Los Alamos National Laboratory under project number 20230363ER.






\appendix

\section{Derivation of Eqs.~\eqref{ETHz-approx}--\eqref{gnsl-def}}
\label{Appx:SaddlePointInt}

We begin our analysis by approximating the sinc-function entering Eq.~\eqref{ETHz-sinc} to second order in the wavevector mismatch, followed by performing the following transformations
\begin{eqnarray}
\label{sinc-approx}
&~&\hspace{-1cm}\frac{\sin\left(\Delta k_{msl}(\omega,-\omega')L_z/2\right)}{\Delta k_{msl}(\omega,-\omega')}
    \approx\frac{L_z}{2i}\left[1-\frac{L^2}{24} \Delta k^2_{msl}(\omega,-\omega') \right]
\\\nonumber&~&
    =\frac{L_z}{2i}\exp\left(\ln\left(1
    -\frac{L^2}{24} \Delta k^2_{msl}(\omega,-\omega') \right)\right)
\\\nonumber&~&
    \approx\frac{L_z}{2i}\exp\left(-\frac{L^2}{24} \Delta k^2_{msl}(\omega,-\omega') \right).   
\end{eqnarray}
Here, the final expression is derived by truncating the expansion of the logarithm in the exponential at second order in the wavevector mismatch. Upon expanding the wavevector mismatch in a Taylor series near the pump central frequency $\omega_0$, 
\begin{eqnarray}
\label{dk-approx}
&~&\hspace{-0.5cm}\Delta k_{msl}(\omega,-\omega_0-\delta\omega)\approx\Delta k_{msl}(\omega,-\omega_0)
\\\nonumber&~&    
    +\partial_{\omega_0}\Delta k_{msl}(\omega,-\omega_0)\delta\omega
    +\partial^2_{\omega_0}\Delta k_{msl}(\omega,-\omega_0)\frac{\delta\omega^2}{2}, 
\end{eqnarray}
we substitute the result into the last line of Eq.~\eqref{sinc-approx}. Keeping only terms up to the second order in $\delta\omega$, we arrive at the following approximate expression for the sinc-function
\begin{eqnarray}
\label{sinc-exp}
&~&\hspace{-1.0cm}\frac{\sin\left(\Delta k_{msl}(\omega,-\omega_0-\delta\omega)L_z/2\right)}
        {\Delta k_{msl}(\omega,-\omega_0-\delta\omega)}
\\\nonumber&~&
    \approx\frac{L_z}{2i}\exp\left(-\frac{L^2}{24} \Delta k^2_{msl}(\omega,-\omega_0)
\right.\\\nonumber&~&\left.      
    -\frac{L_z^2}{12}\Delta k_{msl}(\omega,-\omega_0)\partial_{\omega_0}\Delta k_{msl}(\omega,-\omega_0)\delta\omega
\right.\\\nonumber&~&\left.  
-\frac{L^2}{24} \left\{\left[\partial_{\omega_0}\Delta k_{msl}(\omega,-\omega_0)\right]^2 
\right.\right.\\\nonumber &~& \left. \left.
   +\Delta k_{msl}(\omega,-\omega_0)\partial^2_{\omega_0}\Delta k_{msl}(\omega,-\omega_0)\right\}\delta\omega^2
    \right).   
\end{eqnarray}

Similarly, we apply the following Taylor expansion to the effective second-order optical susceptibility which was defined in Eq.~\eqref{chi2-eff}
\begin{eqnarray}
\label{chi2-approx}
&~&\hspace{-1cm} \tilde\chi_{msl}^{(2)}(\omega;\omega+\omega_0+\delta\omega,-\omega_0-\delta\omega)
\\\nonumber&~&
    \approx\tilde\chi_{msl}^{(2)}(\omega;\omega+\omega_0,-\omega_0)
\left\{
\right.\\\nonumber&~&\left.
    1+\frac{\partial_{\omega_0}\tilde\chi_{msl}^{(2)}(\omega;\omega+\omega_0,-\omega_0)}
    {\tilde\chi_{msl}^{(2)}(\omega;\omega+\omega_0,-\omega_0)}\delta\omega
\right.\\\nonumber&~&\left. 
+\frac{\partial^2_{\omega_0}\tilde\chi_{msl}^{(2)}(\omega;\omega+\omega_0,-\omega_0)}
    {2\tilde\chi_{msl}^{(2)}(\omega;\omega+\omega_0,-\omega_0)}\delta\omega^2
   \right\}. 
\end{eqnarray}
After exponentiating and expanding the logarithm of the expression within the curly brackets up to second-order terms in $\delta\omega$, we derive
\begin{eqnarray}
\label{chi2-exp}
&~&\hspace{-1cm} \tilde\chi_{msl}^{(2)}(\omega;\omega+\omega_0+\delta\omega,-\omega_0+\delta\omega)
\\\nonumber&~&
\approx\tilde\chi_{msl}^{(2)}(\omega;\omega+\omega_0,-\omega_0)
\\\nonumber&~&\times 
\exp\left(
\frac{\partial_{\omega_0}\tilde\chi_{msl}^{(2)}(\omega;\omega+\omega_0,-\omega_0)}
    {\tilde\chi_{msl}^{(2)}(\omega;\omega+\omega_0,-\omega_0)}\delta\omega
\right.\\\nonumber&~&\left. 
+\left\{\frac{\partial^2_{\omega_0}\tilde\chi_{msl}^{(2)}(\omega;\omega+\omega_0,-\omega_0)}
    {2\tilde\chi_{msl}^{(2)}(\omega;\omega+\omega_0,-\omega_0)}
\right.\right.\\\nonumber&~&\left.\left.
   -\frac{1}{2}\left[\frac{\partial_{\omega_0}\tilde\chi_{msl}^{(2)}(\omega;\omega+\omega_0,-\omega_0)}
    {2\tilde\chi_{msl}^{(2)}(\omega;\omega+\omega_0,-\omega_0)}\right]^2\right\}  \delta\omega^2
   \right). 
\end{eqnarray}

Using Eqs.~\eqref{sinc-exp} and \eqref{chi2-exp}, we express the product of the sinc-function and the second-order susceptibility as
\begin{eqnarray}
\label{chi2-sinc-exp}
&~&\hspace{-0.5cm}\tilde\chi_{msl}^{(2)}(\omega;\omega+\omega_0+\delta\omega,-\omega_0+\delta\omega)
\\\nonumber &~&
\times\frac{\sin\left(\Delta k_{msl}(\omega,\omega_0+\delta\omega)L_z/2\right)}{\Delta k_{msl}(\omega,\omega_0+\delta\omega)}
\\\nonumber &~&\approx\frac{L_z}{2i} \tilde\chi_{msl}^{(2)}(\omega;\omega+\omega_0,-\omega_0)
\\\nonumber &~&
\times\exp\left(-\kappa_{msl}(\omega)-\tau_{msl}(\omega)\delta\omega-\frac{\delta\omega^2}{\sigma^2_{msl}(\omega)}\right),
\end{eqnarray}
where the shorthand notations are defined as
\begin{eqnarray}
\label{kappa-msl-def}
\kappa_{msl}(\omega)&=& \frac{L_z^2}{24}\Delta k_{msl}^2(\omega,-\omega_0), 
\\\label{tau-msl-def}
\tau_{msl}(\omega)&=& \frac{L_z^2}{12}\Delta k_{msl}(\omega,-\omega_0)\partial_{\omega_0}\Delta k_{msl}(\omega,-\omega_0)
\\\nonumber &-&
\frac{\partial_{\omega_0}\tilde\chi_{msl}^{(2)}(\omega;\omega+\omega_0,-\omega_0)}
    {\tilde\chi_{msl}^{(2)}(\omega;\omega+\omega_0,-\omega_0)},
\\ \label{sigma-msl-def}
\sigma^{-2}_{msl}(\omega)&=& 
    \frac{L_z^2}{24}\left\{\left[\partial_{\omega_0}\Delta k_{msl}(\omega,-\omega_0)\right]^2 
\right.\\\nonumber &+& \left.
    \Delta k_{msl}(\omega,-\omega_0)\partial^2_{\omega_0}\Delta k_{msl}(\omega,-\omega_0)\right\}
\\\nonumber &-&
\frac{\partial^2_{\omega_0}\tilde\chi_{msl}^{(2)}(\omega;\omega+\omega_0,-\omega_0)}
    {2\tilde\chi_{msl}^{(2)}(\omega;\omega+\omega_0,-\omega_0)}
\\\nonumber &+&
\frac{1}{2}\left[\frac{\partial_{\omega_0}\tilde\chi_{msl}^{(2)}(\omega;\omega+\omega_0,-\omega_0)}
    {2\tilde\chi_{msl}^{(2)}(\omega;\omega+\omega_0,-\omega_0)}\right]^2.
\end{eqnarray}

Finally, by substituting Eq.~\eqref{chi2-sinc-exp} along with Eq.~\eqref{Ep-Gauss} into Eq.~\eqref{ETHz-sinc}, we transform the expression for the transmitted THz pulse into the following approximate form
\begin{eqnarray}
\label{ETHz-sexp}
&~&\hspace{-1cm}{\cal E}^{(t)}_m(\omega)\approx\frac{L_z\omega}{2n(\omega)c}
\tilde\chi_{msl}^{(2)}(\omega;\omega+\omega_0,-\omega_0){\cal A}_s{\cal A}_l
\\\nonumber &~&\times
   \int d\delta\omega \exp\left(-\frac{\left[\omega+\delta\omega\right]^2+\delta\omega^2}{2\sigma^2_\text{p}}  
\right.\\\nonumber&~&\left. 
-\kappa_{msl}(\omega)-\tau_{msl}(\omega)\delta\omega-\frac{\delta\omega^2}{\sigma^2_{msl}(\omega)}
  \right) . 
\end{eqnarray}
This equation yields Eqs.~\eqref{ETHz-approx} and \eqref{gnsl-def} after a straightforward evaluation of the Gaussian integral.

\bibliography{biblocal}

\end{document}